\pdfoutput=1
\PassOptionsToPackage{pdfpagelabels=false}{hyperref} 
\documentclass[fleqn,usenatbib]{mnras}

\usepackage{pdflscape}

\usepackage{graphicx}	
\usepackage{amsmath}	
\usepackage{amssymb}	
\usepackage{verbatim}
\usepackage{paralist}
\usepackage{tasks}
\usepackage{hyperref}

\title[Optical emission of ULXs above Eddington]{Modelling optical emission of Ultra-luminous X-ray Sources accreting above the Eddington limit}

\author[E. Ambrosi and L. Zampieri]{Elena Ambrosi$^{1,2}$ and Luca Zampieri$^{2}$
\\
$^{1}$Department of Physics and Astronomy, University of Padova, Italy \\
$^{2}$INAF-Astronomical Observatory of Padova, Italy
}

\date{Accepted XXX. Received YYY; in original form ZZZ}

\pubyear{2018}

\begin{document}
\label{firstpage}
\pagerange{\pageref{firstpage}--\pageref{lastpage}}
\maketitle

\begin{abstract}
We study the evolution of binary systems of Ultra-luminous X-ray sources and compute their optical emission assuming accretion onto a black hole via a non standard, advection-dominated slim disc with an outflow. We consider systems with black holes of $20M_{\odot}$ and $100M_{\odot}$, and donor masses between $8M_{\odot}$ and $25M_{\odot}$. Super-critical accretion has considerable effects on the optical emission. The irradiating flux in presence of an outflow remains considerably stronger than that produced by a standard disc. However, at very high accretion rates the contribution of X-ray irradiation becomes progressively less important in comparison with the intrinsic flux emitted from the disc. After Main Sequence the evolutionary tracks of the optical counterpart on the colour-magnitude diagram are markely different from those computed for Eddington-limited accretion. Systems with stellar-mass black holes and $12-20 M_{\odot}$ donors accreting supercritically are characterized by blue colors (F450W -- F555W $\simeq - 0.2 : +0.1$) and high luminosity ($M_{V} \simeq - 4 : - 6.5$). Systems with more massive black holes accreting supercritically from evolved donors of similar mass have comparable colours but can reach $M_V \simeq - 8$. We apply our model to NGC 1313 X-2 and NGC 4559 X-7. Both sources are well represented by a system accreting above Eddington from a massive evolved donor. For NGC 1313 X-2 the agreement is for a $\sim 20M_{\odot}$ black hole, while NGC4559 X-7 requires a significantly more massive black hole.
\end{abstract}


%
\section{Introduction}
Ultra-luminous X-ray Sources (ULXs) are non nuclear point-like extragalactic X-ray sources with bolometric luminosity higher than the Eddington limit for a 10$M_{\odot}$ black hole ($L_{Edd} \sim 2 \times 10^{39}$ erg s$^{-1}$; \citealt{1989ARA&A..27...87F}). The majority of them are X-ray binaries with a neutron star (NS) or a black hole (BH), most probably accreting above the Eddington limit. In the past years, there has been an extensive debate on the possibility that the compact object is a stellar or a massive stellar BH accreting above Eddington (\citealt{2009MNRAS.400..677Z,2009MNRAS.395L..71M,2010ApJ...714.1217B,2011NewAR..55..166F}), or an Intermediate mass Black Hole accreting sub-Eddington \citep{1999ApJ...519...89C}. The recent discovery of pulsating ULXs (\citealt{2014Natur.514..202B,2016ApJ...831L..14F,2017Sci...355..817I,2017MNRAS.466L..48I}) led to the conclusion that the accretor can in fact also be a NS.

In the last 15 years a significant number of stellar optical counterparts of ULXs have been identified and progressively investigated in detail by several authors (e.g. \citealt{2002ApJ...580L..31L,2004ApJ...603..523Z,2005MNRAS.356...12S,
2007ApJ...658..999M,2008A&A...486..151G,2011ApJ...734...23G}). \cite{2011ApJ...737...81T} and \cite{2013ApJS..206...14G} performed an homogeneous reanalysis of all the available {\it Hubble Space Telescope} ({\it HST}) photometric data of stellar counterparts of ULXs. In the majority of the cases, they appear to be hosted in young stellar environments (e.g. \citealt{2006ApJ...641..241R,2006IAUS..230..293P,2007ApJ...661..165L,2008A&A...486..151G,2011ApJ...734...23G}) and have magnitudes and colours consistent with those of O-B type stars. As for Galactic X-ray binaries, the optical emission of ULX binaries originates from the donor star and the outer regions of the accretion disc. The X-ray flux produced in the innermost regions of the disc can intercept the outer regions and be thermalized, resulting in an enhancement of the optical-IR flux (e.g. \citealt{1993PASJ...45..443S} for standard accretion onto a Schwarzschild BH).
The peculiar character of ULX binaries is that X-ray irradiation could be dominant also for massive donors.\\

Comparison of stellar evolutionary tracks of ULXs with the photometric properties of their optical counterparts on the colour-magnitude diagram may be used to constrain the masses of their donor stars (e.g. \citealt{2005MNRAS.356...12S,2005MNRAS.362...79C,2007MNRAS.376.1407C}). \cite{2008MNRAS.386..543P,2010MNRAS.403L..69P} (hereafter PZ and PZ2, respectively) produced evolutionary tracks of ULXs binary systems accreting onto a BH via a standard self-irradiated accretion disc. They showed that a massive donor is needed to fuel persistent ULXs at the required rates (see also \citealt{2005MNRAS.364..344P,2005MNRAS.356..401R}) and provided constrains on both the donor and the BH. Several ULXs are also associated with very extended optical emission nebulae, that give important information on their energetics and lifetime \citep{2002astro.ph..2488P,2003MNRAS.342..709R}.

In this work we consider a ULX with a stellar or massive-stellar BH accreting super-Eddington. This configuration can explain some basic facts concerning the X-ray spectral components and the short-term variability at high energies observed in some ULXs (\citealt{2015MNRAS.447.3243M}) and is supported also by the detection of emission lines and blueshifted absorption lines from highly ionized Fe, O and Ne in high-resolution X-ray spectra (\citealt{2016Natur.533...64P}). Super-critical accretion was considered already in \cite{1973A&A....24..337S}, who proposed that in such conditions radiation pressure could originate an outflow from the innermost regions of the accretion disc. \cite{1988ApJ...332..646A} first studied in detail the properties of such optically thick, advection-dominated accretion discs (slim discs). More recently, supercritical accretion onto BHs has been studied by means of 2D radiation and magneto-hydro simulations, that show the formation of an advection-dominated disc and an outflow region, with powerful clumpy winds driven by radiation pressure (e.g. \citealt{2009PASJ...61..783T,2011ApJ...736....2O,2013PASJ...65...88T}).

In the following we study the evolution of ULXs binary systems and compute their optical emission assuming that they accrete onto a BH via a non standard, advection-dominated slim disc with an outflow. The plan of the paper is the following. In Section 2 we summarize the model of PZ for Eddington-limited accretion in ULX binaries, which is at the base of this work. In Section 3 we describe the implementation of our model for super-Eddington accretion in ULX binaries. In Section 4 we show the evolutionary tracks of our systems in the color-magnitude diagram and analyze the results obtained with and without an outflow. We then show a preliminary application of our model to NGC 1313 X-2 and NGC 4559 X-7. Finally, in Section 5 we discuss our results and further developments of this work.

\section{Self-irradiated accretion discs in ULX binaries: the PZ model}

The starting point of our calculation is the model of PZ, that includes X-ray irradiation. As in PZ, we compute the evolution of representative ULX binary systems with black holes (BHs) of 20 and 100$M_{\odot}$, and donor masses in the range between 8-25 $M_{\odot}$. The system is evolved up to the giant phase by means of an updated version of the Eggleton code. The emission properties and the track followed on the colour-magnitude diagram (CMD) are computed with a dedicated code during the contact phases. We assume Population I chemical composition (Helium abundance Y = 0.28, metal abundance Z = 0.02), a mixing length parameter $\alpha = 2.0$ and an overshoot constant $\delta_{ov}=1.2$. Wind mass loss from the donor has been taken into account following the prescription in \citealt{1988A&AS...72..259D} (see \citealt{2005MNRAS.364..344P}, \citealt{2006ApJ...640..918M} and PZ for details). However, accretion takes place only when the donor fills its Roche Lobe.

In PZ accretion is assumed to be always sub-Eddington and the accretion rate is not allowed to exceed the critical rate $\dot{M}_{crit} \simeq 10 \dot{M}_{Edd}$ (for a standard efficiency of 0.1; $\dot{M}_{Edd} = L_{Edd}/c^2$). Once the mass transfer rate sets in, a standard accretion disc (\cite{1973A&A....24..337S}) forms. Its inner and outer radii are assumed to be equal to $r_{in} = 6 r_{g}$, where $r_{g}$ is the gravitational radius, and $r_{out} = 2 r_{circ}$, where $r_{circ}$ is the circularization radius.\footnote{The circularization radius $r_{circ}$ is defined as the radius where the Keplerian angular momentum is equal to the angular momentum of the gas streaming from the inner Lagrangian point. The accreting gas settles initially at $r_{circ}$ and then spreads at smaller and larger radii because of viscous dissipation and conservation of angular momentum, forming a disc (see e.g. \citealt{frank2002accretion}, chapter 5.2 for details). Typically, $r_{circ}$ is a significant fraction of the Roche lobe radius, so that the disc outer boundary $r_{out}$ cannot extend more than a few times $r_{circ}$. We take $r_{out} = 2 r_{circ}$ as a reference value.} A fraction of the X-ray flux emitted from the inner regions of the disc is absorbed and thermalized in the outer parts. Following \cite{2005MNRAS.362...79C}, the local temperature stratification induced by such X-ray heating is given by:
\begin{equation}
\centering
T(\tau) = \Bigl\{\frac{\pi}{\sigma}\Bigl[B_{x}(\tau)+B_{d}(\tau)\Bigr]\Bigr\}^{1/4} \, .
\label{eq:temp_strat}
\end{equation}
Here $B_{x}(\tau)$ is the energy flux originating from X-ray irradiation, while $B_{d}(\tau)$ refers to the contribution of viscous heating. The former takes the form:
\small
\begin{eqnarray}\label{eq:bbirr}
&&B_{x}(\tau)=\frac{1}{2} S_{x}\biggl\{ k_{s}f_{s}(\alpha)\left(\frac{\xi}{1+\xi}\right)\left[A_{s}-\left(A_{s}-\frac{1}{2}\right)e^{\tau/A_{s}}\right] \nonumber \\
 && +k_{h}f_{h}(\alpha)\left(\frac{1}{1+\xi}\right)\left[A_{h}+\left(A_{h}-\frac{1}{2}\right)e^{\tau/A_{h}}\right]\biggr\}.
\end{eqnarray}
\normalsize
where $S_{x}$ is the incident X-ray flux absorbed by an annulus of the disc, $\tau$ is the optical depth to optical photons, $\xi = S_{h}/S_{s}$ is the hardness parameter (ratio of the hard and soft X-ray flux), $k_{s}$ and $k_{h}$ are the opacities (in units of the electron scattering opacity) for the soft and hard components of the incident radiation, $f_{s}(\alpha)$ and $f_{h}(\alpha)$ are functions that depend on the geometry of the system ($\alpha$ being the angle of incidence of the X-rays measured from the normal to the surface), $A_{s} = cos(\alpha)/k_{s}$ and $A_{h} = cos(\alpha)/k_{h}$.
The incident X-ray flux has been calculated from the following expression:
\begin{equation}
S_{x}= \frac{L_x}{(1-A)4\pi r^2}
\label{eq:irr_stand}
\end{equation}
where $L_{x} = \eta \dot{M}c^2$ is the bolometric luminosity of the source and $A$ is the albedo to X-rays, fixed at $A = 0.9$\footnote{Varying the value of the albedo $A$ between 0.7 and 0.95 has no major effects on the position and shape of the evolutionary tracks on the CMD, as shown in PZ2.}. The disc spectrum is calculated integrating over the disc annuli and assuming that they emit as black bodies at the temperature given by equation~(\ref{eq:temp_strat}) (where we set $\tau = 2/3$ as the photospheric optical depth).

Concerning the star, assuming it at a distance equal to the binary separation $\textit a$, its temperature is given by an expression similar to equation~(\ref{eq:temp_strat}):
\begin{equation}
T_{star} = \Bigl[T^4_{unirr,star}+T^4_{irr,star}\Bigr]^{1/4} \, ,
\label{eq:temp_stratstar}
\end{equation}
where $\sigma T^4_{unirr,star}$ and $\sigma T^4_{irr,star}$ represent the energy fluxes of the unirradiated star and that originating from X-ray irradiation, respectively. The latter is calculated from:
\begin{equation}
T_{irr,star} = \Bigl[\frac{L_{x}(1-A)}{8 \pi \sigma a^2 }\Bigr] \, .
\label{eq:irrstar}
\end{equation}
For further details on the adopted irradiation model we refer to PZ, \cite{2001MNRAS.320..177W}, and \cite{2005MNRAS.362...79C}.

\subsection{Photometry of the donor}\label{subs:phot}

In order to perform a direct comparison of the emission properties of ULX binary systems with the available {\it HST} photometric measurements, in the present version of the code we implemented a different method for calculating the magnitudes of the donor stars.
While PZ perform an interpolation of the tabulated magnitudes and bolometric corrections (BC) reported in \cite{cox1999editor}, here we use the code of \cite{2003MmSAI..74..474G} to calculate the donor magnitudes for a given luminosity, mass and metallicity. Thus, output magnitudes of the donor are now given in different photometric systems, including the Johnson and {\it HST} systems.

\section{Modelling super-Eddington accretion in ULX binaries}

In the present model, when the accretion rate is below the Eddington limit, the treatment of the accretion flow follows that of PZ. However, in ULX binaries the mass transfer rate from the donor can exceed the Eddington rate. When $\dot{M}\geq \dot{M}_{crit}$, accretion can no longer proceed via a standard thin disc and a bimodal structure is assumed: an inner disc with non-standard disc geometry and temperature profile (vertical scale height comparable to the radius and $T \propto r^{-1/2}$), and an outer disc with standard structure. In addition, a massive optically thick outflow (with an inner advection dominated region) is assumed to set in. In these physical conditions, also the geometry of X-ray irradiation varies.
\cite{1988ApJ...332..646A} investigated accretion at $\dot{M}\geq \dot{M}_{crit}$ assuming a vertically integrated disc structure. In these discs (usually referred to as slim discs) the radial infall velocity and the horizontal pressure gradient become important and can no longer be neglected in the Euler equation. The radial infall velocity of the gas is comparable to the rotational velocity. As the optical depth in the disc is very large, the radiation diffusion timescale becomes larger than the infall timescale and the gas transports a large fraction of the internal energy inwards, producing a strong horizontal heat flux. Such an advective cooling changes the energy balance of the disc. As a consequence, the disc scale height and the radial temperature profile differ significantly from the standard case. Moreover, a large fraction of the advected heat is accreted by the BH and the disc luminosity is no longer proportional to the accretion rate, but tends asymptotically to $\approx 10 L_{Edd}$ \citep{2000PASJ...52..133W}.

When the luminosity exceeds the Eddington limit, the huge pressure force produced by the emitted X-ray flux effectively removes gas from the surface of the disc, overcoming the gravitational force and leading to the formation of a massive radiation-driven outflow. This effect is clearly seen in recent 2D magneto-hydro simulations of super-Eddington accretion (e.g \citealt{2013PASJ...65...88T}). Observational evidence of the existence of high velocity outflows in ULXs comes from the discovery of emission and absorption lines in their optical (\citealt{2015NatPh..11..551F}) and X-ray \citep{2016arXiv161100623P} spectra. Then, in the following we assume that, when the mass transfer rate exceeds the Eddington limit, a bimodal disc structure sets in: an outer standard disc and an inner advection-dominated slim disc with an outflow. The radius where the transition between the two regimes occurs, as well as the onset and extension of the outflow, depend on the mass transfer rate. 

\subsection{Bimodal disc without outflow}\label{ssec:advection_no_outflow}

\begin{figure}
 \includegraphics[width= \columnwidth]{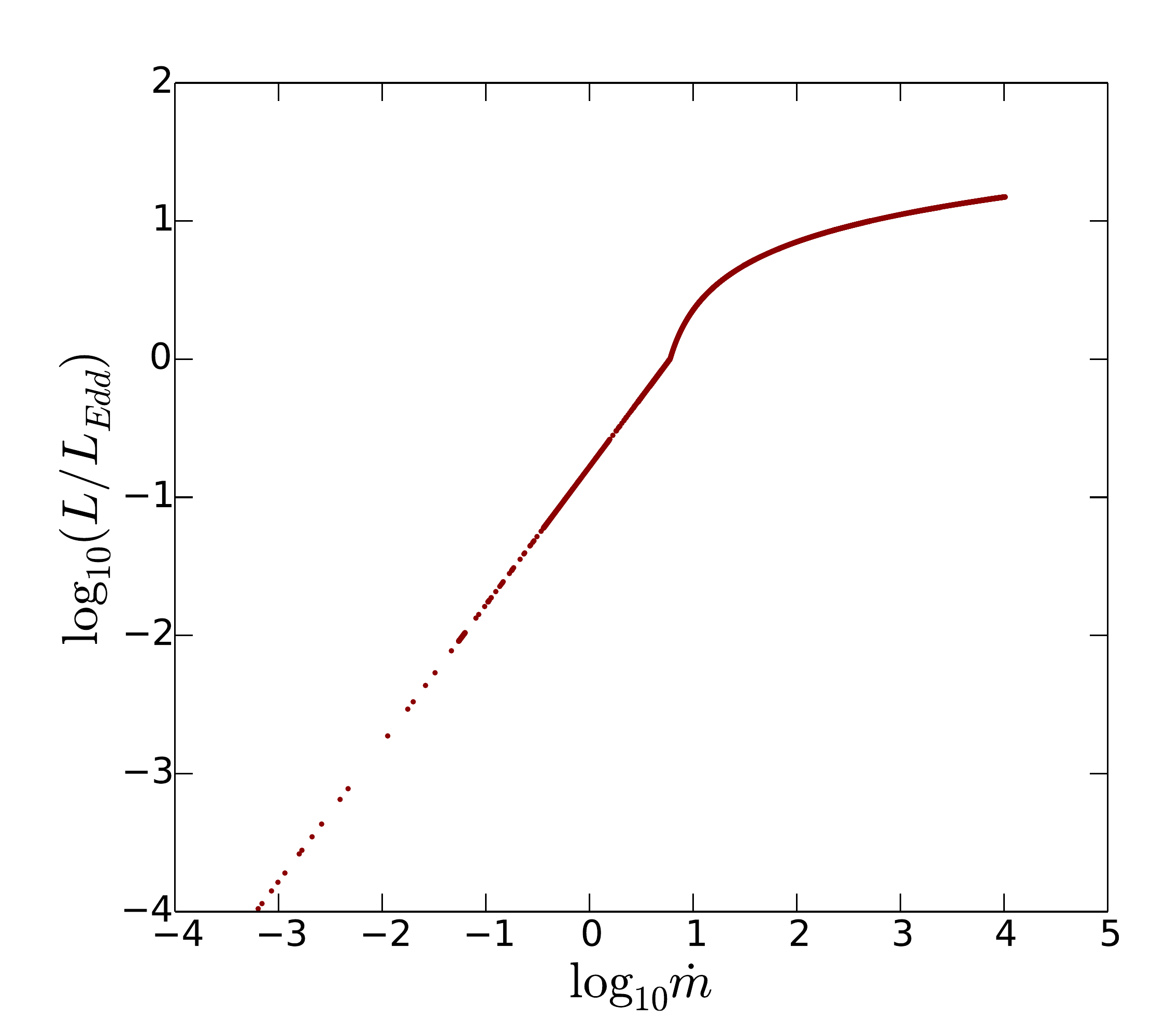}
 \caption{Luminosity of a bimodal disc without outflow as a function of the mass transfer rate. The curve is obtained plotting the luminosity calculated at different phases and for different systems (donors of 10 and 25 $M_{\odot}$, BHs of 20 and 100 $M_{\odot}$).
 }
 \label{fig:lum}
\end{figure}

As a starting point towards a full implementation of our model, we first consider the case of an hybrid (inner advection-dominated plus outer viscosity-dominated) disc structure without an outflow. The innermost and outermost radii of the disc are $r_{in} = 6 r_{g}$ and $r_{out} = 2 r_{circ}$. For the inner slim disc we adopt the parameterization (for self-similar solutions) of \cite{2000PASJ...52..133W}. They define the boundary radius $r_{0}$ as the radial coordinate where the structure of the disc changes (the inward advected heat is equal to the viscously dissipated heat). It is:
\begin{equation}
r_{0} \propto \dot{m} r_{g} \, ,
\label{eq:r0}
\end{equation}
where $\dot{m}=\dot{M}/\dot{M}_{Edd}$. As $\dot{m}$ increases, the inner slim disc becomes more and more extended. This bimodal configuration affects the temperature profile (and hence the emitted spectrum) of the disc: in the inner region, where advection dominates, the temperature varies as $T(r) \propto r^{-1/2}$, while in the outer region, where viscous heating is dominant, $T(r) \propto r^{-3/4}$. Fig.~\ref{fig:lum} shows the luminosity calculated integrating the emitted flux over all the annuli of the disc and assuming local blackbody emission. When advection becomes important, the disc luminosity significantly overcomes the Eddington limit but it never exceeds $\approx 10 L_{Edd}$ (\citealt{1999PASJ...51..725W,2000PASJ...52..133W}). In fact, as $\dot m$ 
increases, the photon mean free path becomes progressively smaller and the radiation diffusion timescale progressively larger. As a consequence, more and more internal energy is effectively advected inward on a dynamical timescale and not radiated away.

The geometry of the inner slim portion of the disc significantly affects also self-irradiation. The flux emitted from the inner regions intercepts the outer parts if $dH/dr \ll 1$ and $H/r \ll 1$ (see e.g. \citealt{frank2002accretion}). In the inner regions of a slim disc $H \simeq r$ and therefore the X-ray flux produced inside is not able to illuminate the outside. In these conditions the geometry of disc self-irradiation must be reconsidered.

\begin{figure}
	\centering
	\includegraphics[width =\columnwidth]{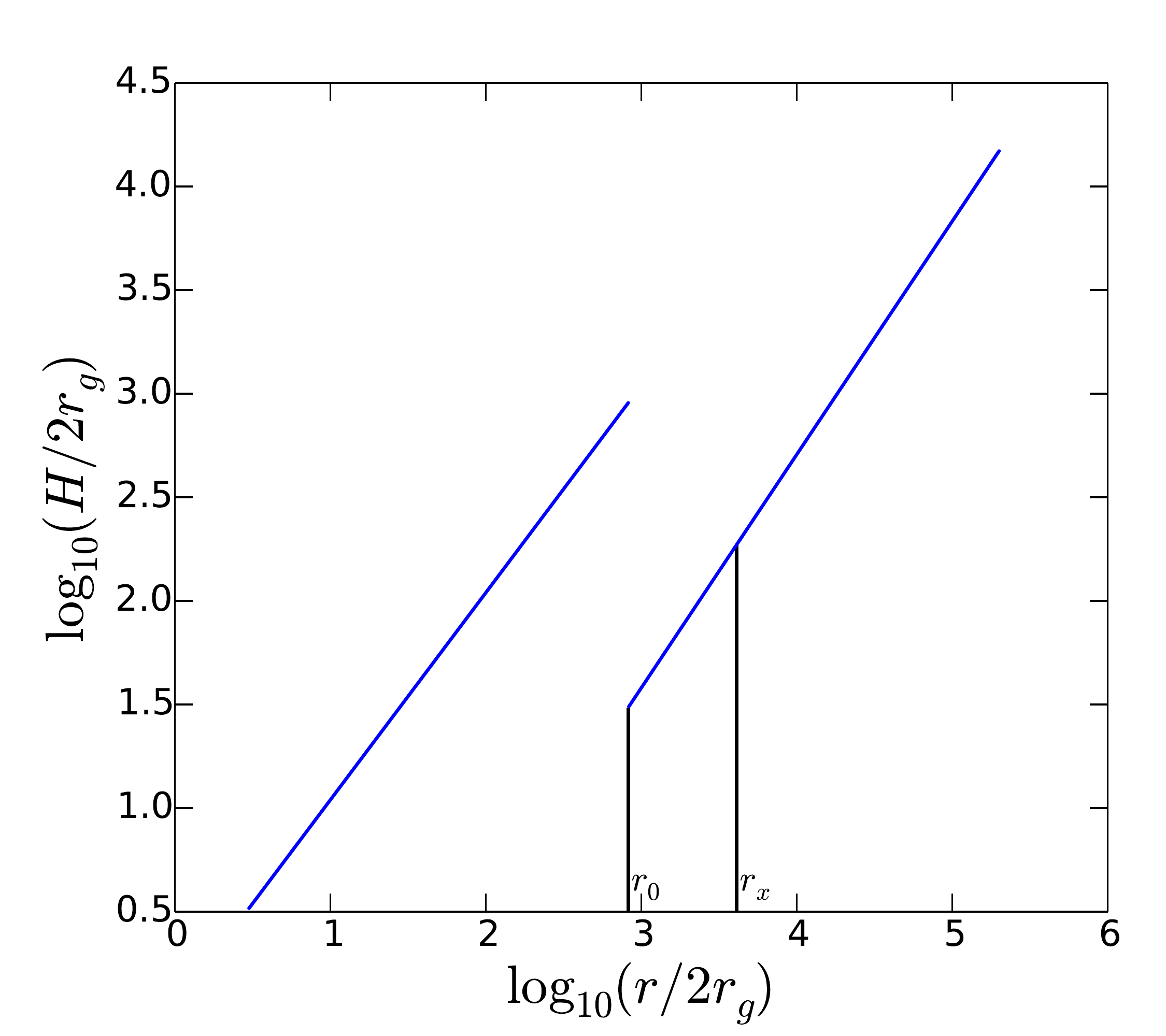}
	\caption{Inner slim disc and outer standard disc scale height (in units of $2 r_g$ and in logaritmic scale) in our bimodal-disc model ($\dot m = 509$). The portion of the standard disc that can irradiate the outer regions is that between $r_0$ and $r_x$.}
	\label{fig:xrayregion}
\end{figure}

\subsubsection{Geometry of disc self-irradiation}

\begin{figure}
\includegraphics[width = \columnwidth]{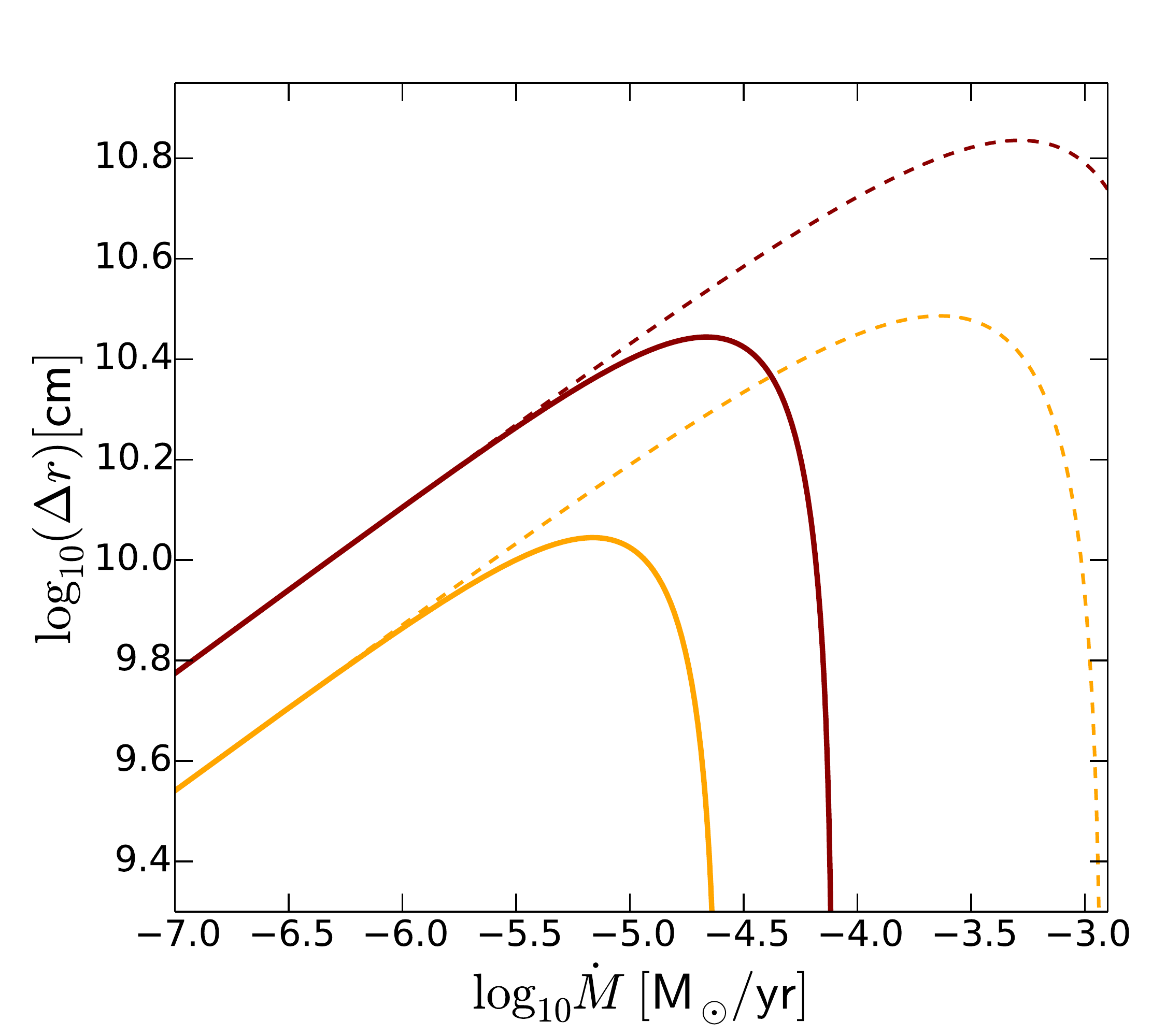}
\caption{X-ray-UV irradiating region of the disc for the model without outflow ({\it dashed} line) and with outflow ({\it solid} line). The ({\bf red}) curves on the top refer to a BH of $100 M_{\odot}$f, while the ({\bf orange}) curves on the bottom to a BH of $20 M_{\odot}$. For the case without outflow $\Delta r = r_{x}-r_{0}$, while for the case with outflow $\Delta r = r_{x}-r_{ph}$.}
 \label{fig:irr_region}
\end{figure}

In our hybrid disc structure the X-ray irradiating flux is produced outside the boundary radius $r_{0}$. In these conditions, the fact that X-ray emission from the inner disc is isotropic or partially beamed has no effect on the outside region. We consider as irradiating flux all the radiation emitted between radii $r_0$ and $r_{x}$, where $r_{x}$ marks approximately the boundary of the region where the gas temperature falls within the X-rays-ultraviolet (UV) energy band (see Fig.~\ref{fig:xrayregion}). If the temperature at $r_{0}$ is lower than $\simeq 5 \times 10^5$ K, no X-ray irradiation takes place. The extension of the irradiating region ($\Delta r = r_{x}-r_{0}$) for BHs of 20 $M_\odot$ and 100 $M_\odot$ is shown in Fig.~\ref{fig:irr_region}. The extension of the irradiating region depends on the mass transfer rate, since both $r_{0}$ and $r_{x}$ depend on $\dot m$. $\Delta r$ reaches a maximum at around a few thousands ${\dot m}$ and then decreases sharply. The irradiation luminosity shows a similar trend with the mass transfer rate.
Clearly, the irradiating region is less extended than in the standard case. However, the X-ray-UV irradiation luminosity is bigger, because the flux increases with the mass transfer rate.

As the results of the evolutionary tracks with a bimodal disc structure without an outflow are likely not to be directly relevant for modelling ULXs, they are reported in Appendix ~\ref{appendix:app} and are no longer discussed further.

\begin{figure}
 \includegraphics[width = \columnwidth]{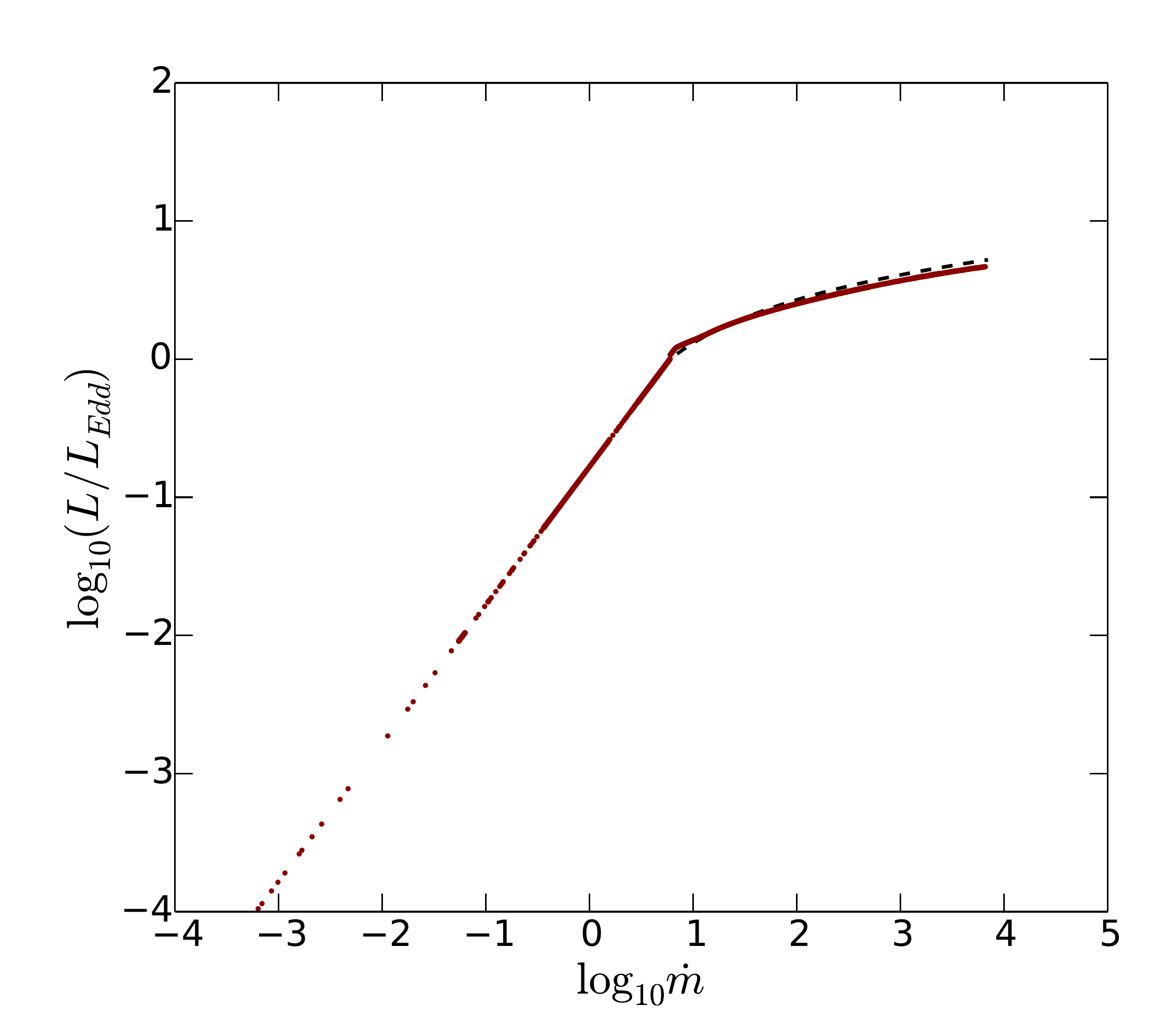}
 \caption{Luminosity of a bimodal disc with outflow as a function of the accretion rate. The {\it dashed} line is the expression reported in equation~(\ref{eq:ltotp}), while the {\it dotted} lines are the values computed for our models (donor mass of 10 $M_\odot$ and 20 $M_\odot$, BH masses 20 $M_\odot$ and 100 $M_\odot$).}
 \label{fig:lpout}
\end{figure}

\subsection{Bimodal disc with outflow}

In addition to a bimodal disc, we included in our model of ULX binaries also an optically thick outflow. For sufficiently large $\dot m$ the outflow can cover all the slim disc and part of the standard disc.
Following the treatment of \cite{2007MNRAS.377.1187P} (to which we refer for a detailed discussion), the outflow can be divided in four zones, depending on the optical depth (measured in the direction perpendicular to the disc plane) and on the mechanism involved in the energy transport. They are delimited by three characteristic radii: $r_{ph,in}$ and $r_{ph,out}$ mark the inner and outer radial locations of the photosphere, while $r_{sph}$ (spherization radius; \citealt{1973A&A....24..337S}) is the radius below (above) which energy transport is dominated by the outward advected heat (radiation diffusion). The expression of these radii is given by \citep{2007MNRAS.377.1187P}:
\begin{eqnarray}
 \frac{r_{ph,in}}{r_{in}} & \approx & 1 + \frac{1}{3\epsilon_{w}} \\
 \frac{r_{sph}}{r_{in}} & \approx & \left[ 1.34 - 0.4\epsilon _{w} + 0.1\epsilon ^{2}_{w} \right. \nonumber \\
                        && \left. - (1.1 - 0.7 \epsilon _{w}) \dot{m_{0}}^{-2/3} \right] \dot{m_{0}} \\
 \frac{r_{ph,out}}{r_{in}} & \approx & 3 \epsilon _{w} \dot{m}_{0}^{3/2} \, ,
 \label{eq:routflow}
\end{eqnarray}
where $\dot{m}_{0}=\dot{M}/\dot{M}_{crit}$ and $\epsilon_{w}$ is the fraction of radiative energy spent in accelerating the outflow. Following \cite{2007MNRAS.377.1187P}, we set $\epsilon_{w} = 0.5$.
%
In terms of these radii the outflow is divided in four zones: 
\begin{itemize}
 \item zone A is the inner optically thin part of the outflow at $r < r_{ph,in}$;
 \item zone B is the inner optically thick part of the outflow between $r_{ph,in}$ and $r_{sph}$, where outward advection of energy dominates;
 \item zone C is the outer optically thick part of the outflow between $r_{sph}$ and $r_{ph,out}$, where diffusion of radiation dominates;
 \item zone D is the outer optically thin part of the outflow at $r > r_{ph,out}$.
\end{itemize}
%
We approximate the temperature profile of the outflow in regions B and C with the following expressions:
\begin{eqnarray}
 && T_{B}(r) = T_{r_{sph}}\bigg(\frac{r}{r_{sph}}\bigg)^{-1/2} \\
 && T_{C}(r) = T_{r_{sph}}\bigg(\frac{r}{r_{sph}}\bigg)^{-3/4} \, ,
 \label{eq:toutflow}
\end{eqnarray}
where $T_{r_{sph}} = 1.5m^{-1/4}\dot{m_{0}}^{-1/2}(1+0.3\dot{m_{0}}^{-3/4})(1-\epsilon _{w}) \, {\rm keV}$ is the temperature at the spherization radius. The total luminosity of the system is calculated integrating the emitted flux over all the annuli of the disc for $r_{in} < r < r_{ph,in}$ and $r_{ph,out} < r < r_{out}$, and over all radial slices of the outflow for $r_{ph,in} < r < r_{ph,out}$. Local blackbody emission at the appropriate temperature is assumed. The final expression of the emitted luminosity is \citep{2007MNRAS.377.1187P}:
\begin{equation}
	L \approx L_{Edd} \left( 1+\frac{3}{5}\ln \dot{m_{0}} \right) \, .
	\label{eq:ltotp}
\end{equation}
This luminosity is shown in Fig.~\ref{fig:lpout}, along with that computed from our models (see below). The fractional difference is at most $10\%$.



\begin{table}
\caption{Donor and BH masses of the binary systems considered in this work}\label{tab:systems}
\begin{center}
\begin{tabular}{|c|c|}
 \hline
 Donor star ($M_{\odot}$) & Black Hole ($M_{\odot}$)\\
 \hline
 8.0; 10.0; 12.0 & 20.0\\
 15.0; 20.0; 25.0 & 20.0\\
 8.0; 10.0; 12.0 & 100.0\\
 15.0; 20.0; 25.0   & 100.0\\
\hline
\end{tabular}
\end{center}
\end{table}

\begin{figure*}
 \includegraphics[width=\columnwidth]{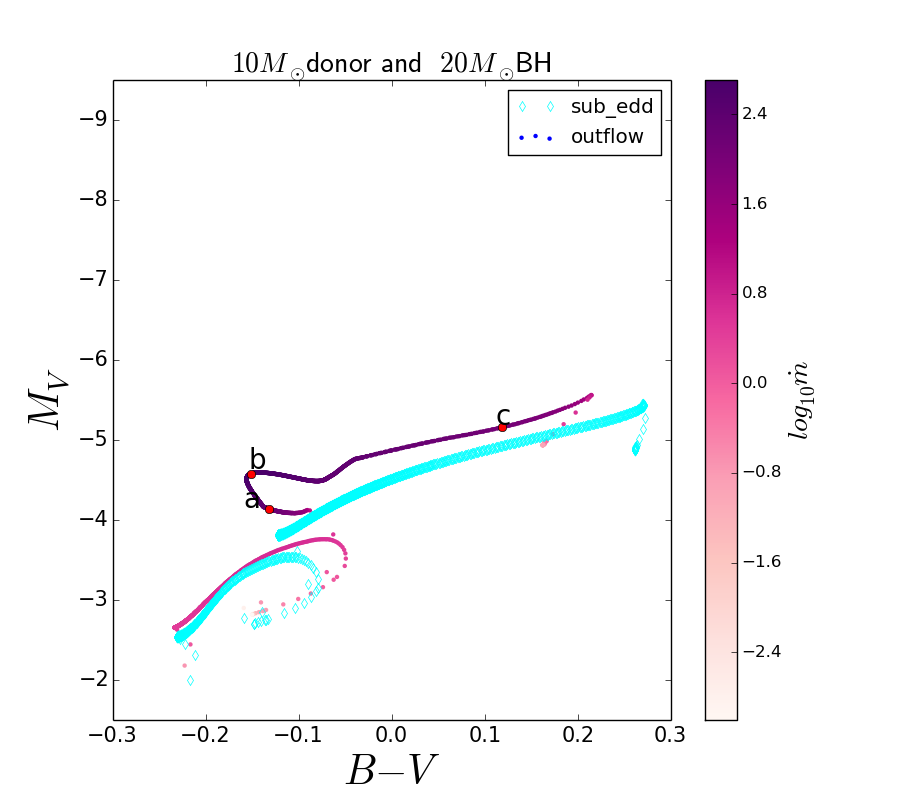}
 \includegraphics[width=\columnwidth]{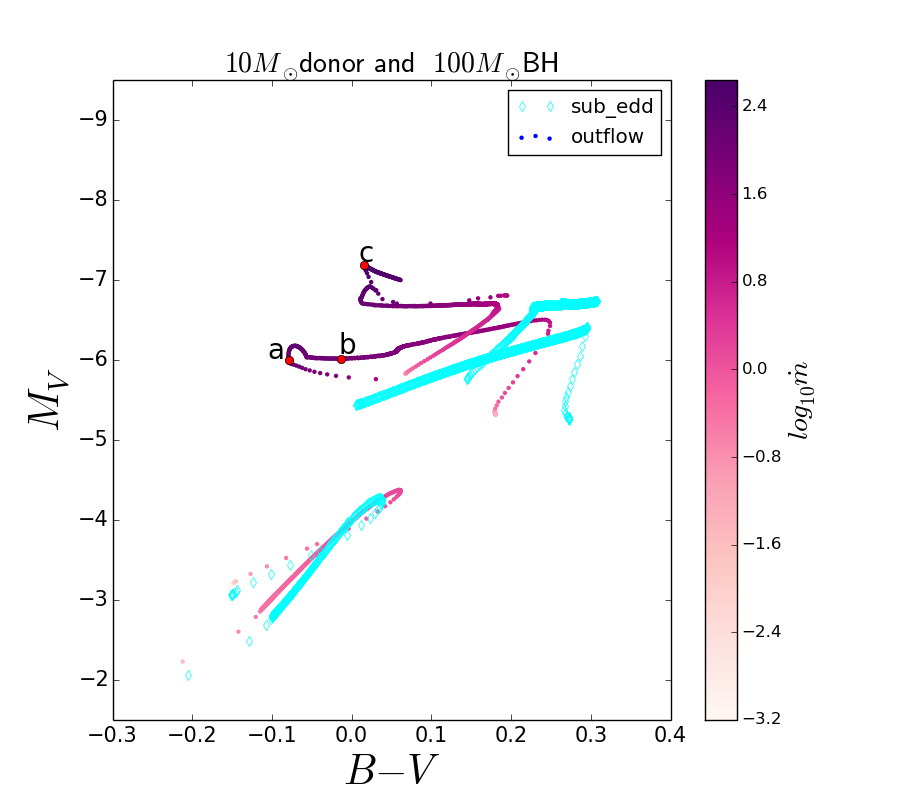}
 \includegraphics[width=\columnwidth]{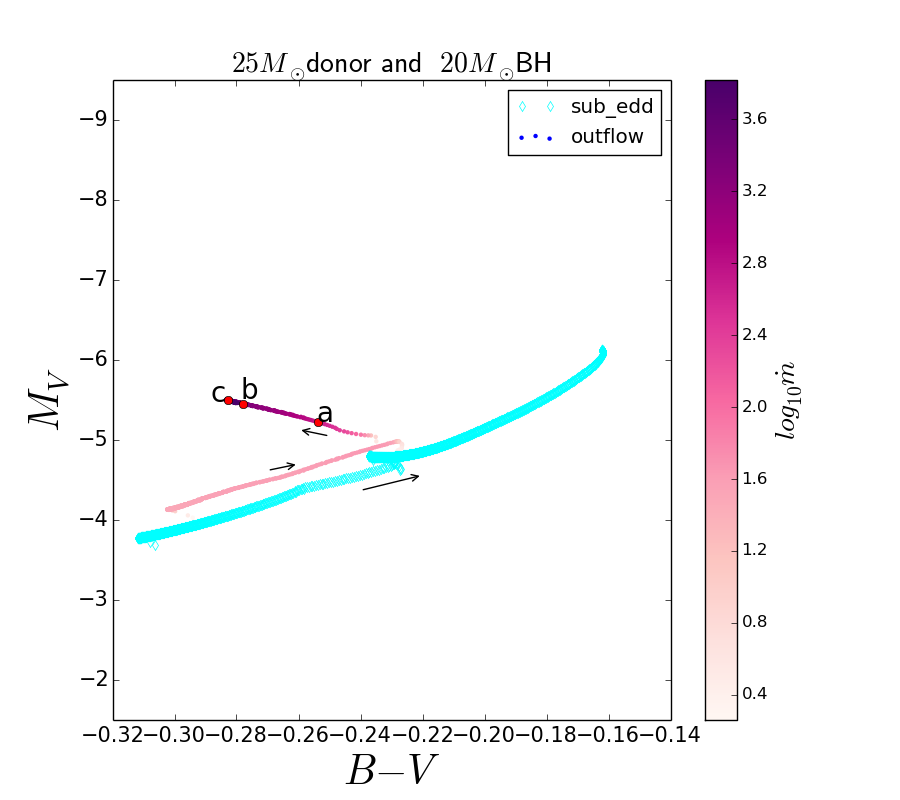}
 \includegraphics[width=\columnwidth]{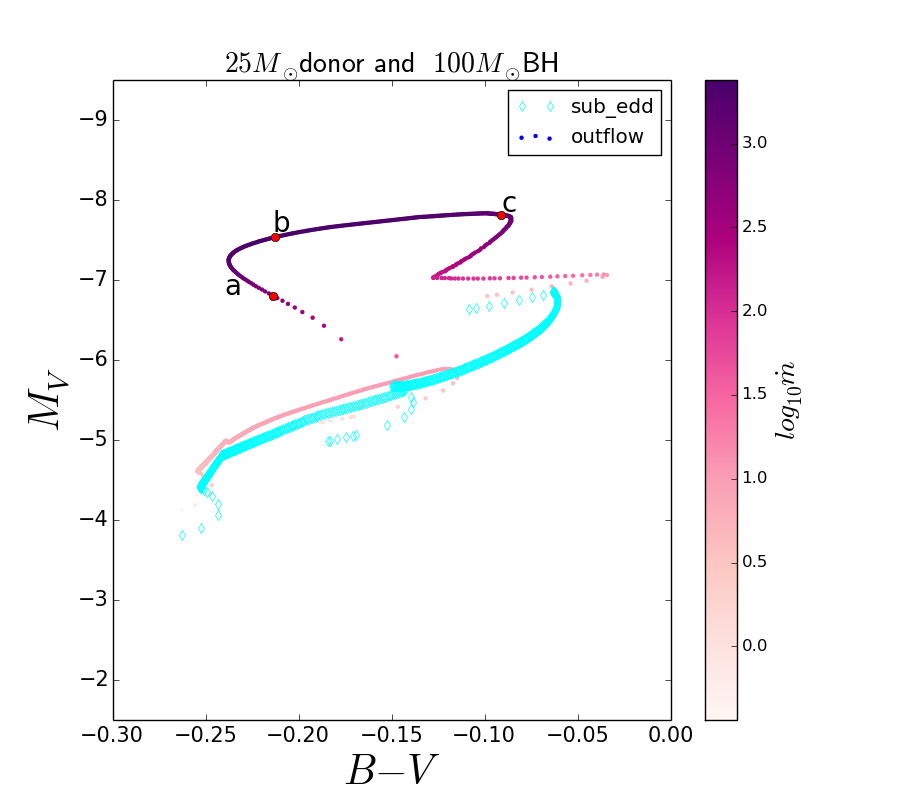}
 %
 \caption{Evolution on the colour-magnitude diagram of a ULX binary system with a donor mass of $10 M_{\odot}$ ({\it upper} panels) or $25 M_{\odot}$ ({\it lower} panels) at zero age main sequence and a BH mass of $20 M_{\odot}$ ({\it left} panels) or $100 M_{\odot}$ ({\it right} panels). The {\it thick} ({\it light grey}, {\it cyan} in the on-line version) line represents the evolution calculated assuming standard sub-Eddington accretion (PZ). The {\it thin} ({\it gray-scale}, {\it color-scale} in the on-line version) line represents the new evolution computed for super-Eddington accretion. Points $a$, $b$ and $c$ mark the evolutionary phases at which the spectral energy distribution is computed. The track of the system with a $25 M_{\odot}$ donor and a $20 M_{\odot}$BH was interrupted when $r_{ph,out}>r_{out}$. Representative arrows in the bottom left panel indicate the direction of the evolution along the tracks.}
 \label{fig:diagrammi*}
\end{figure*}

\subsubsection{Geometry of disc self-irradiation}

The geometry of disc self-irradiation with the outflow is different from that without the outflow. The outflow reprocesses entirely the disc emission up to $r_{ph,out} \gg r_{0}$ and has a scale height of order unity. Consequently, radiation from the disc emitted inside $r_{ph,out}$ cannot hit the region of the disc outside it. The irradiating region is then $\Delta r = r_{x}-r_{ph,out}$, with $r_{ph,out} \gg r_{0}$, and is thus significantly less extended than in the pure slim disc case. Moreover, its dependence on the mass transfer rate is different, as shown in Fig.~\ref{fig:irr_region}. It has a maximum at a few hundreds times $\dot{m}$. An additional contribution to the emitting region may come also from the photosphere of the outflow located at $r_{ph,out}$ and facing the outer disc. We add its contribution to the irradiating flux if its temperature falls in the X-ray-UV band (eq.[\ref{eq:toutflow}]). However, this is usually not sufficient to compensate the 'missing' flux from the region between $r_{0}$ and $r_{ph,out}$, and therefore irradiation is smaller than in the pure slim disc case. Also in this case, the fact that X-ray emission from the inner disc is isotropic or partially beamed has no effect on the outside region.



\section{Results}

\subsection{Photometry}\label{ssec:resphot}

We compared the B, V and R (Johnson) magnitudes obtained with the new treatment of the photometry of the donor star presented in Sec.~\ref{subs:phot} with those of PZ. The accretion disc and X-ray irradiation are not taken into account, while the decrement of the mass of the donor during its evolution is considered. 
The maximum difference between the magnitudes computed with the old and new treatment of photometry occurs for massive ($> 20 M_\odot$) stars during main sequence and is $\sim 40\%$. For comparison, in PZ the maximum fractional deviation of the adopted fitting formula from the tabulated reference magnitudes is $\sim 15\%$. 

\subsection{Evolutionary tracks on the color-magnitude diagram}\label{ssec:cmd_outflow}

We used the model described in the previous Sections to compute the evolution of 12 binary systems with the initial donor and BH masses reported in Tab. ~\ref{tab:systems}. In the following, we will show the results for systems with donors of $10 M_{\odot}$ and $25 M_{\odot}$, and BHs of $20 M_{\odot}$ and $100 M_{\odot}$.
The initial binary separation of all systems is chosen so that the first contact phase occurs during MS. The possibility that accretion starts only after MS is not considered because, for massive donors, this phase is too short to be consistent with the characteristic age of the optical emission nebulae typically associated to ULXs ($>500$ thousand years; \citealt{2002astro.ph..2488P}).

\begin{figure*}
\bfseries
\begin{tasks}[counter-format = {(tsk[a])},label-offset = {0.8em},label-format = {\bfseries}](3)
\task
\includegraphics[width = 0.3\textwidth]{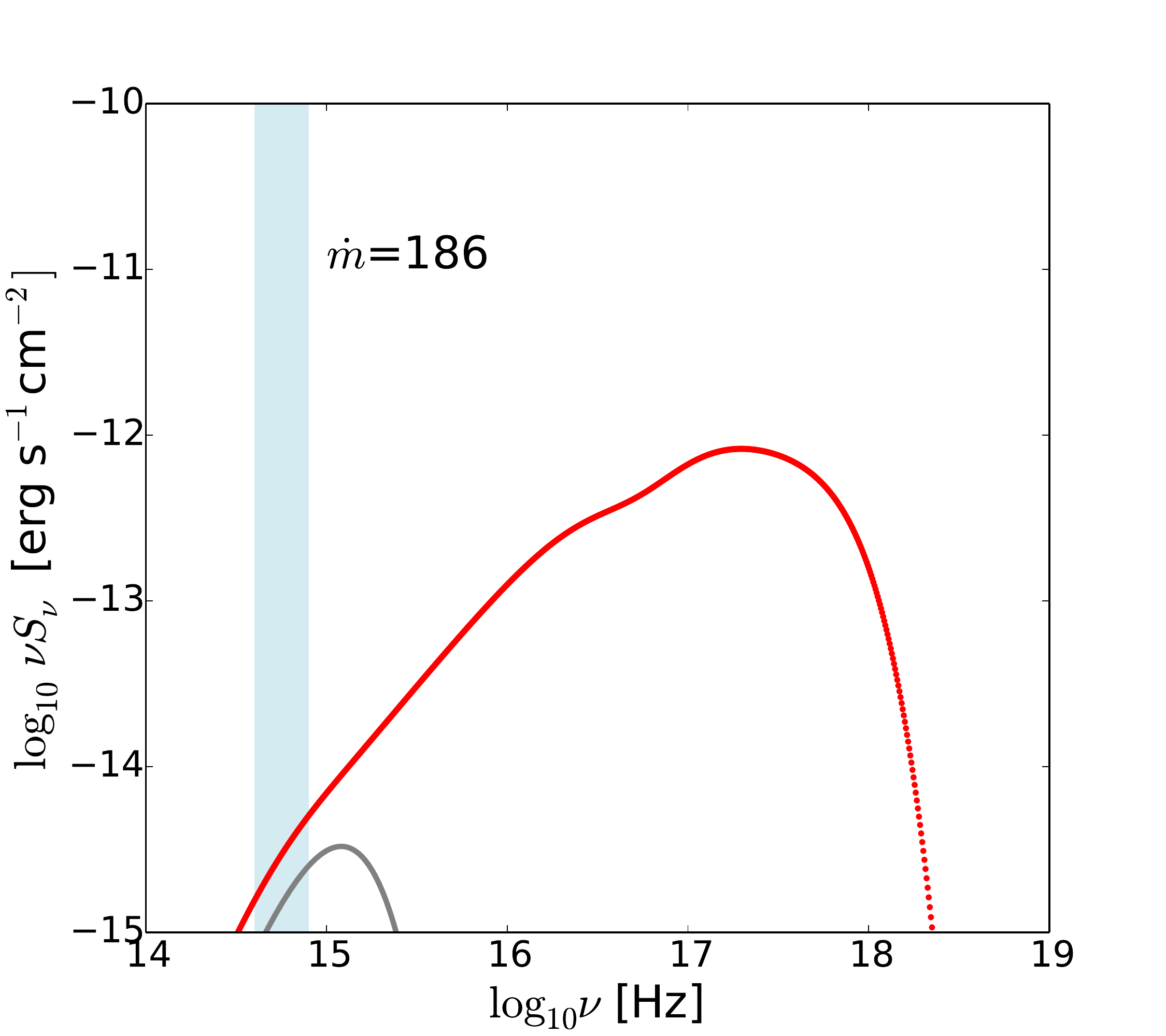}
\task
\includegraphics[width = 0.3\textwidth]{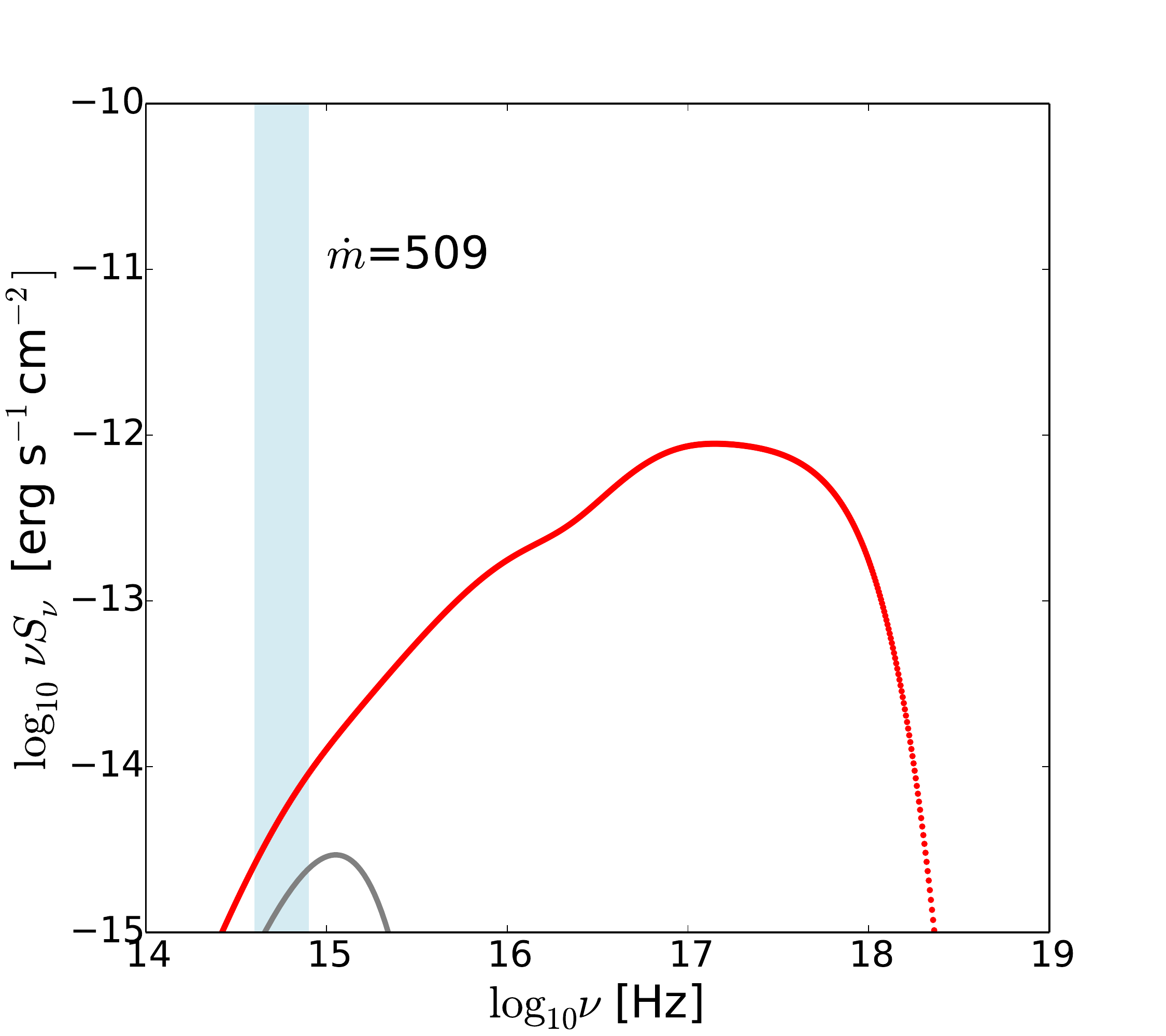}
\task
\includegraphics[width = 0.3\textwidth]{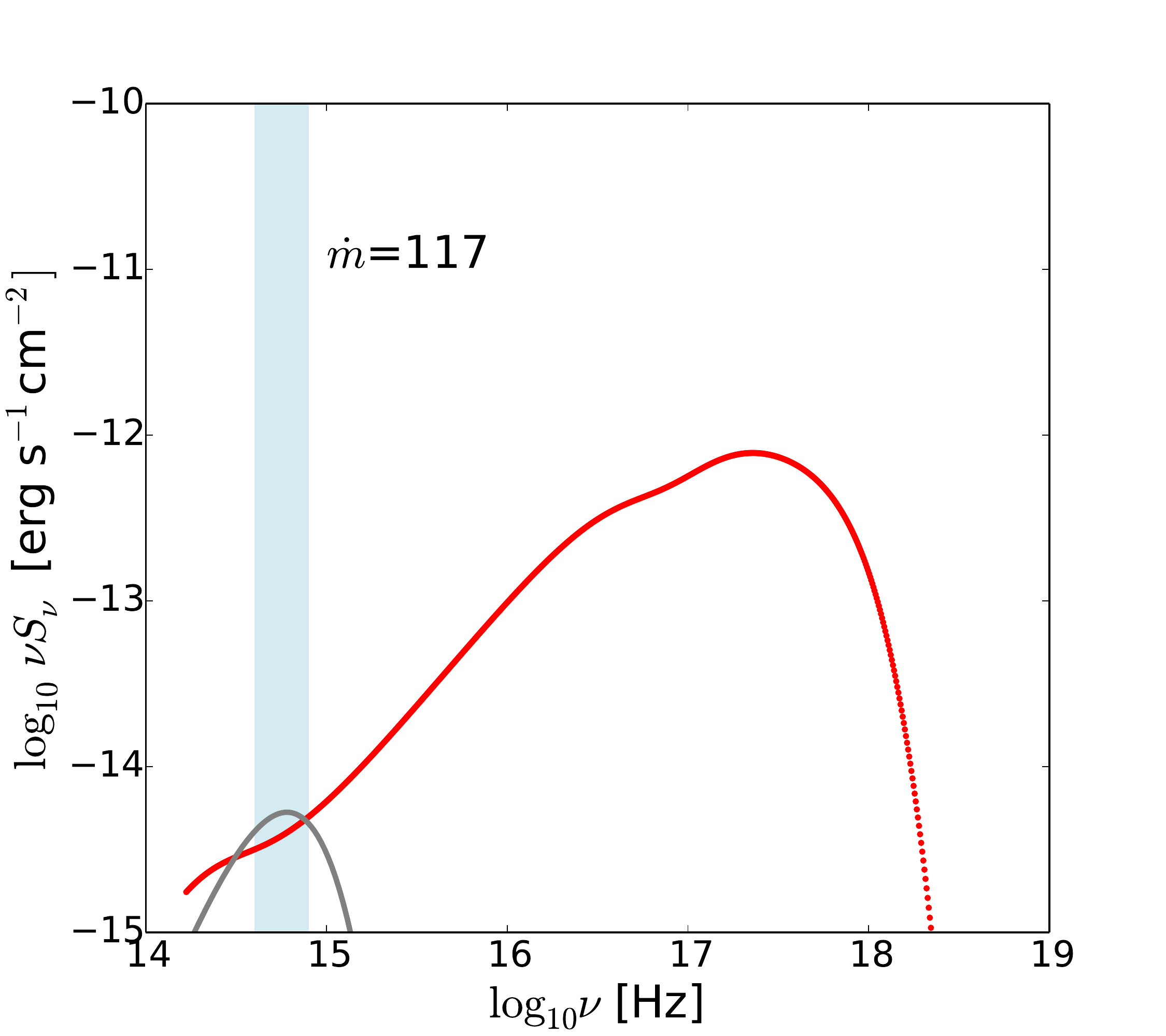}
\end{tasks}
\caption{Optical-through-X-ray spectrum of the system with a $10 M_{\odot}$ donor and a $20 M_{\odot}$ BH at the phases marked with $a$, $b$ and $c$ on the evolutionary track for super-Eddington accretion. The {\it thick} ({\it red}) line represents the spectrum of the self-irradiated outer disc plus the outflow and the innermost slim disc, while the {\it thin} ({\it gray}) line is the spectrum of the X-ray heated donor. The {\it gray} ({\it light blue}) strip marks the optical band. At phases $a$ and $c$ both the disc and the outflow irradiate the outer standard disc, while at phase $b$ only the outflow does it. The mass of the donor star at each time is: $M_{donor,a} \simeq 4.4 M_{\odot}$, $M_{donor,b} \simeq 4.2 M_{\odot}$ and $M_{donor,c} \simeq 2.03 M_{\odot}$.}\label{fig:spettro1020*}
\end{figure*}

\begin{figure*}
\begin{tasks}[counter-format = {(tsk[a])},label-offset = {0.8em},label-format = {\bfseries}](3)
\task
\includegraphics[width = 0.3\textwidth]{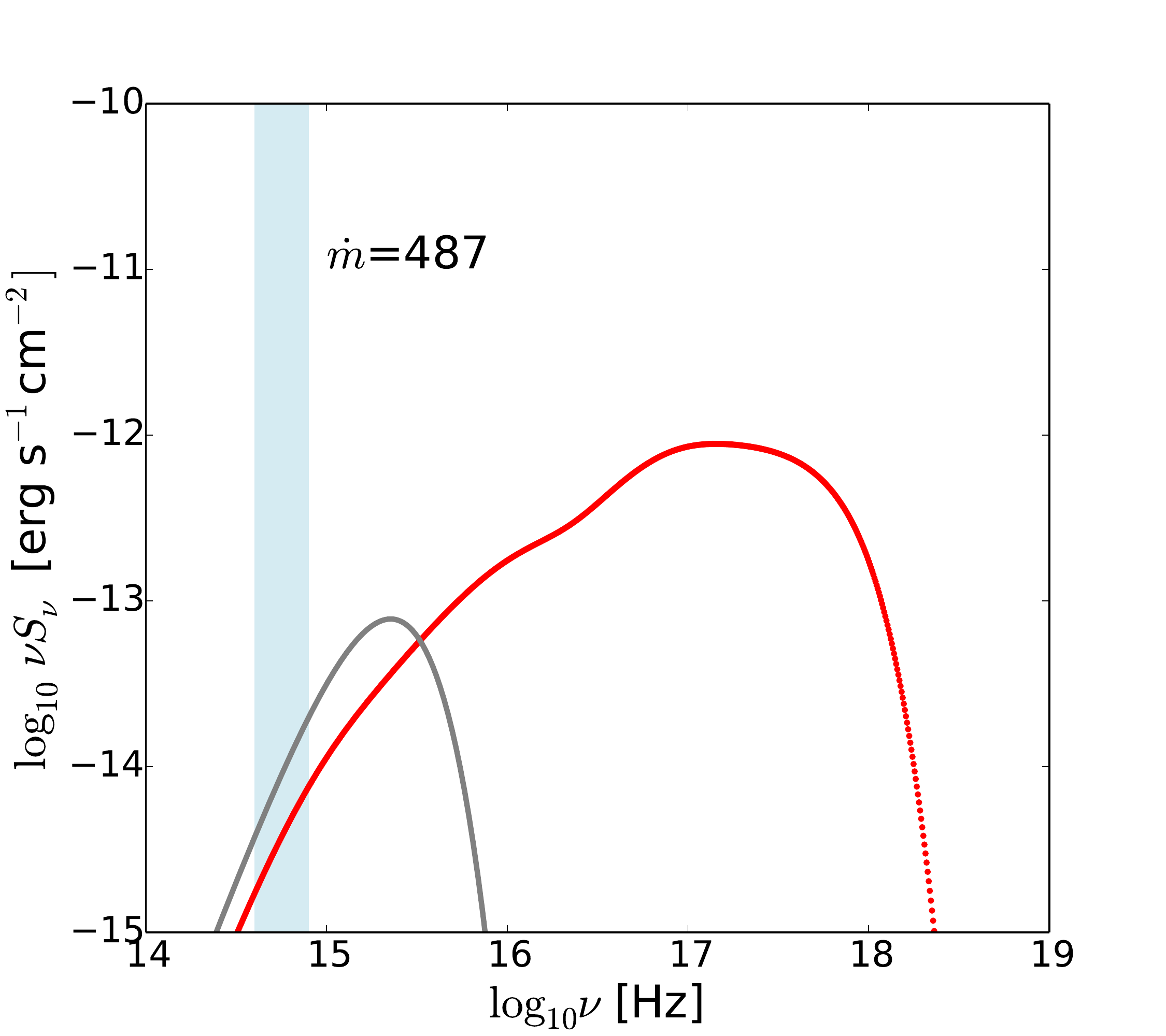}
\task
\includegraphics[width = 0.3\textwidth]{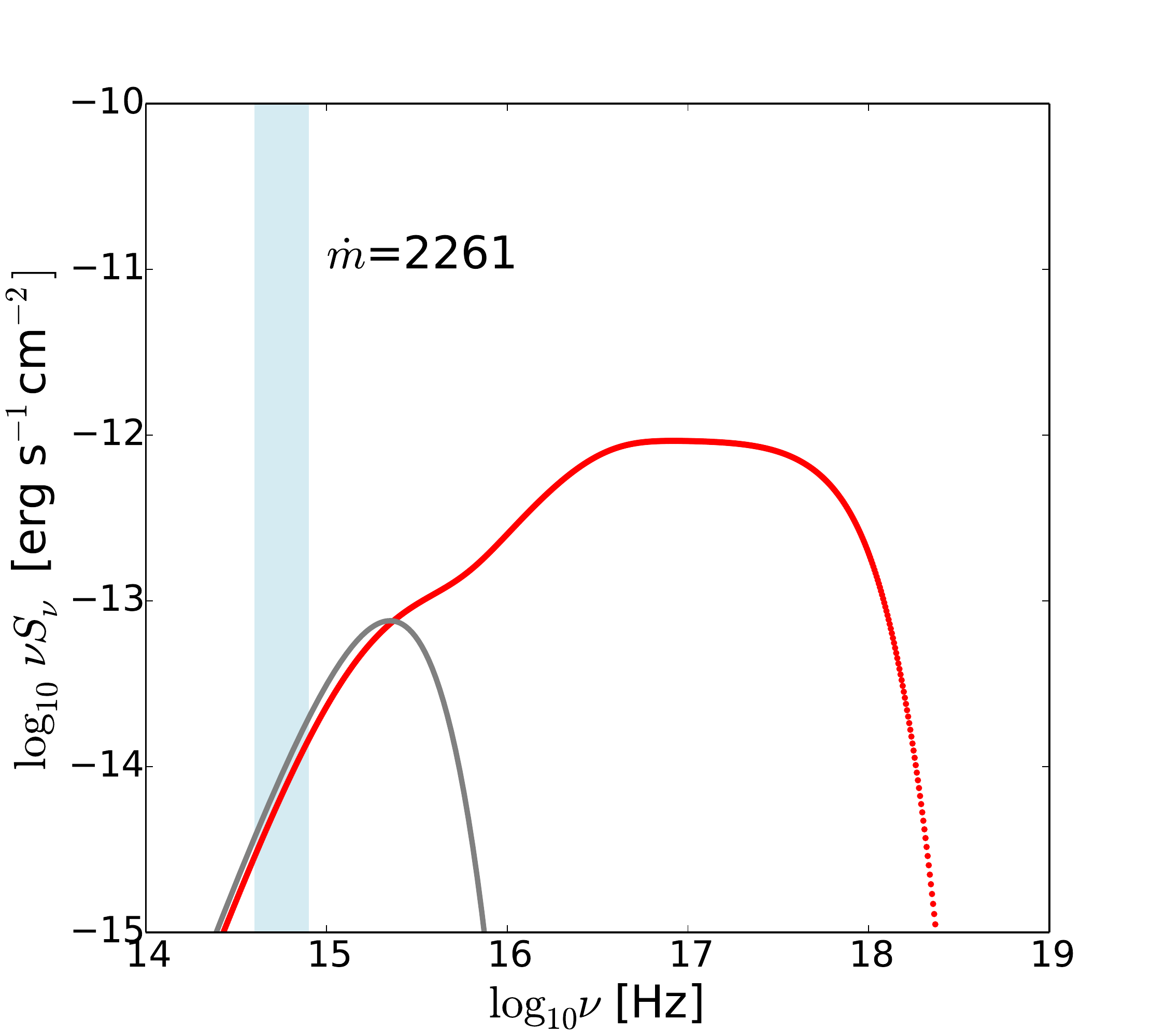}
\task
\includegraphics[width = 0.3\textwidth]{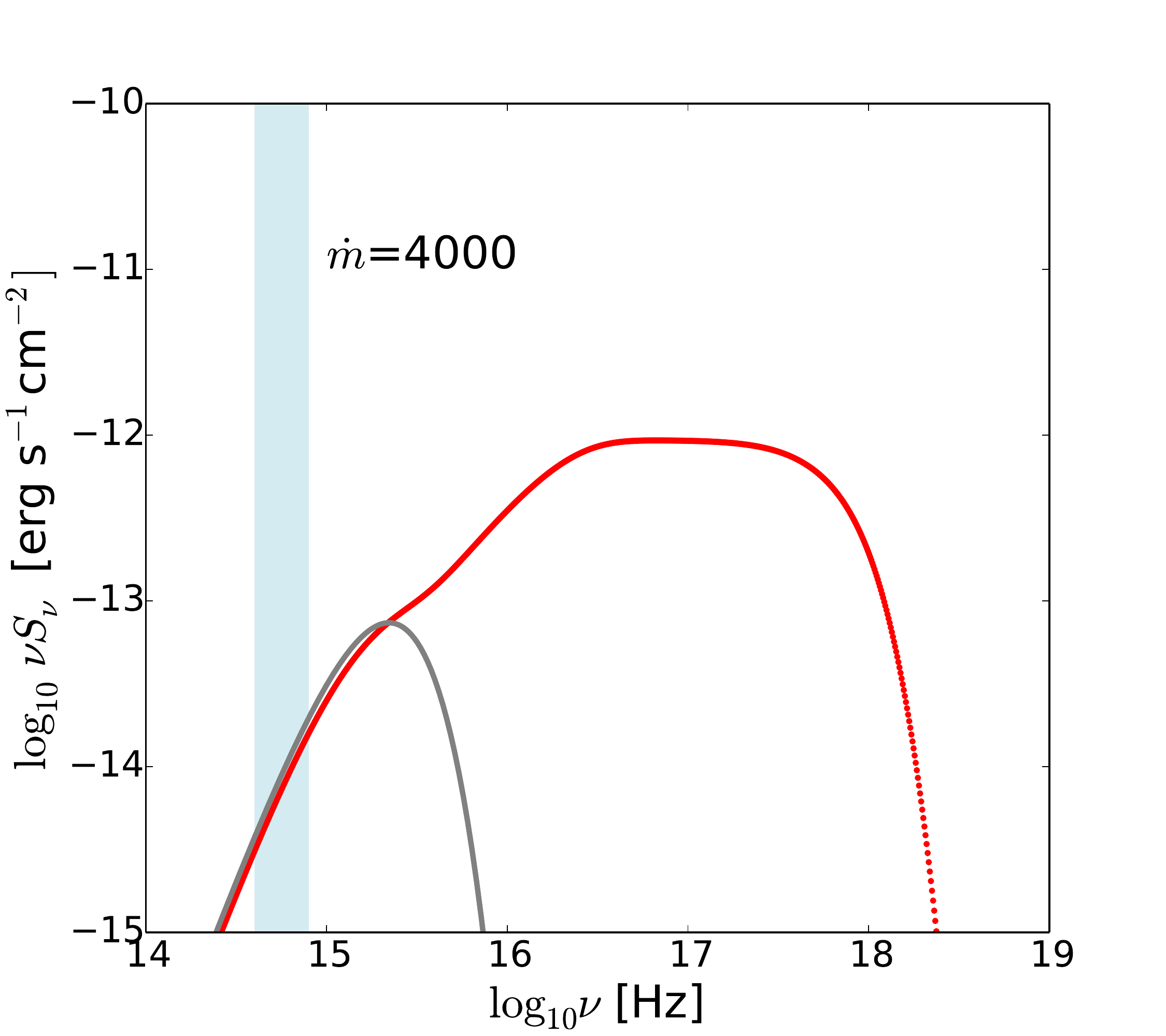}
\end{tasks}
\caption{Same as Figure~\ref{fig:spettro1020*} for a $25 M_{\odot}$ donor and a $20 M_{\odot}$ BH. At phase $a$ the outflow irradiate the outer standard disc, while at phases $b$ and $c$ there is no irradiation. The mass of the donor star at each time is: $M_{donor,a} \simeq 11.80 M_{\odot}$, $M_{donor,b} \simeq 11.78 M_{\odot}$ and $M_{donor,c} \simeq 11.70 M_{\odot}$.}
\label{fig:spettro2520*}
\end{figure*}

\begin{figure*}
\begin{tasks}[counter-format = {(tsk[a])},label-offset = {0.8em},label-format = {\bfseries}](3)
\task
\includegraphics[width = 0.3\textwidth]{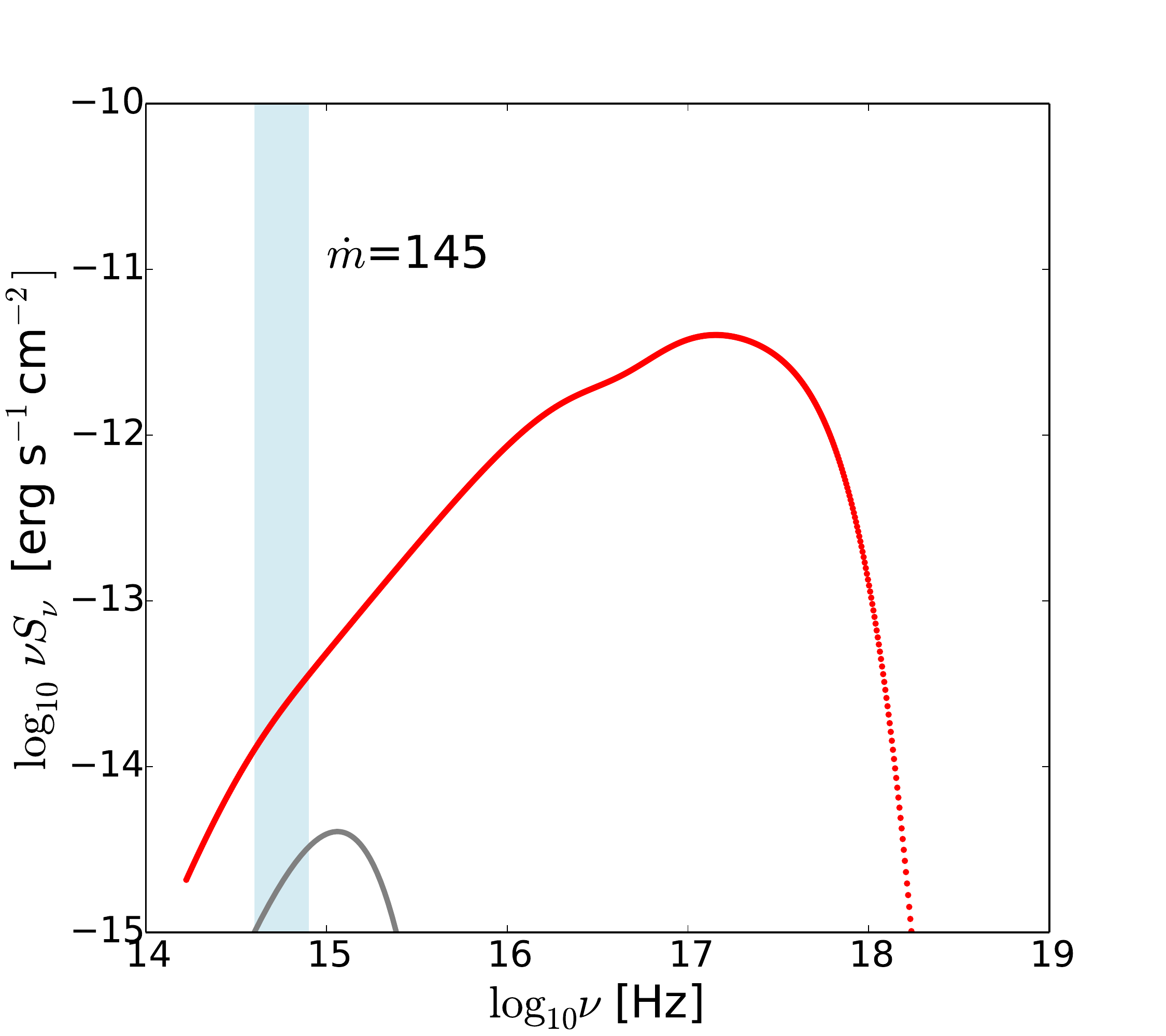}
\task
\includegraphics[width = 0.3\textwidth]{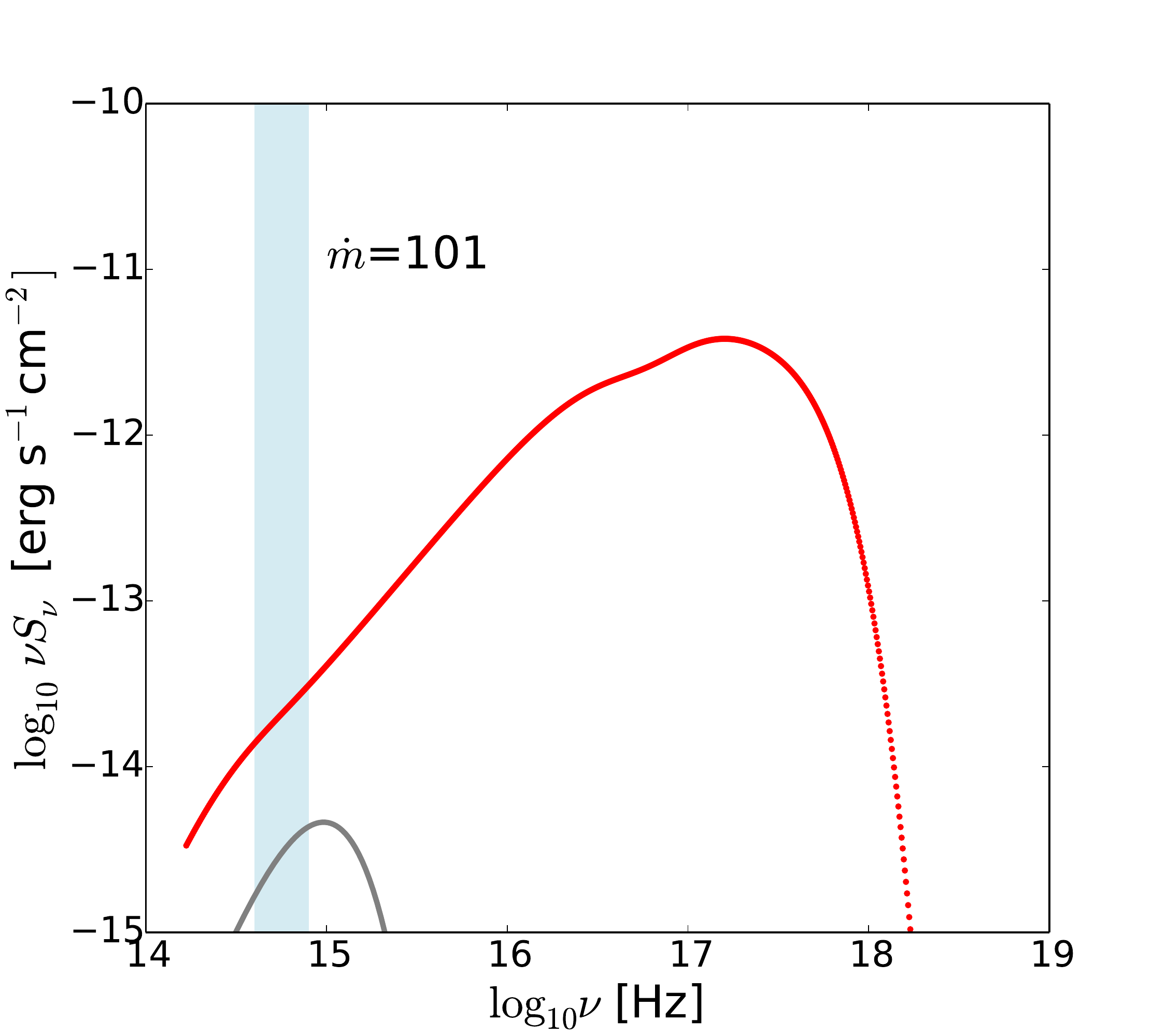}
\task
\includegraphics[width = 0.3\textwidth]{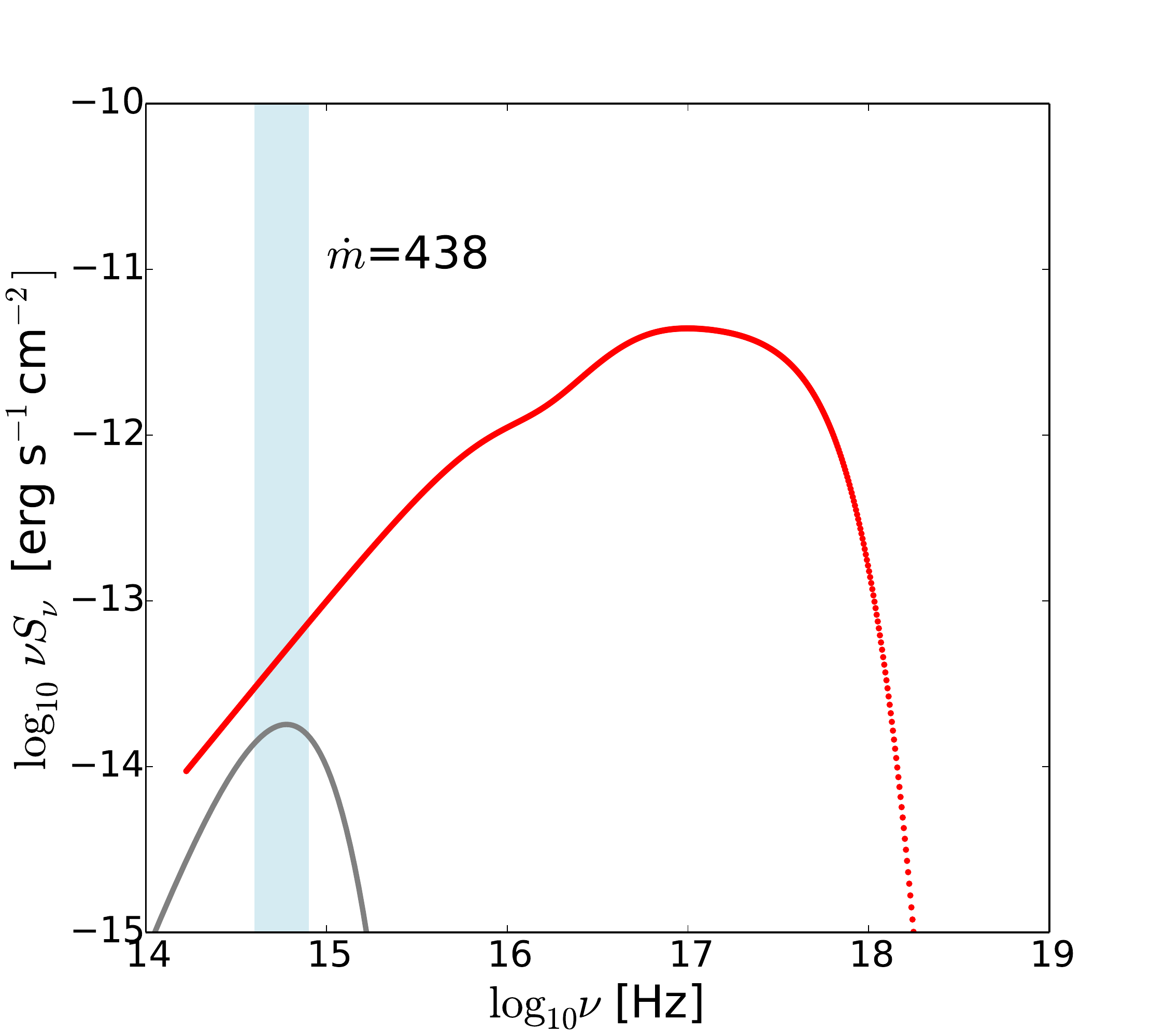}
\end{tasks}
\caption{Same as Figure~\ref{fig:spettro1020*} for a $10 M_{\odot}$ donor and a $100 M_{\odot}$ BH. At phases $a$ and $c$ only the photosphere of the outflow at $r_{ph,out}$ radiates the outer standard disc. In $b$ both the outflow and the disc irradiate the outer standard disc. In $c$ the donor is on the Giant branch. The mass of the donor star at each time is: $M_{donor,a} \simeq 4.99 M_{\odot}$, $M_{donor,b} \simeq 3.90 M_{\odot}$ and $M_{donor,c} \simeq 1.38 M_{\odot}$.}\label{fig:spettro10100*}
\end{figure*}
\begin{figure*}
\begin{tasks}[counter-format = {(tsk[a])},label-offset = {0.8em},label-format = {\bfseries}](3)
\task
\includegraphics[width = 0.3\textwidth]{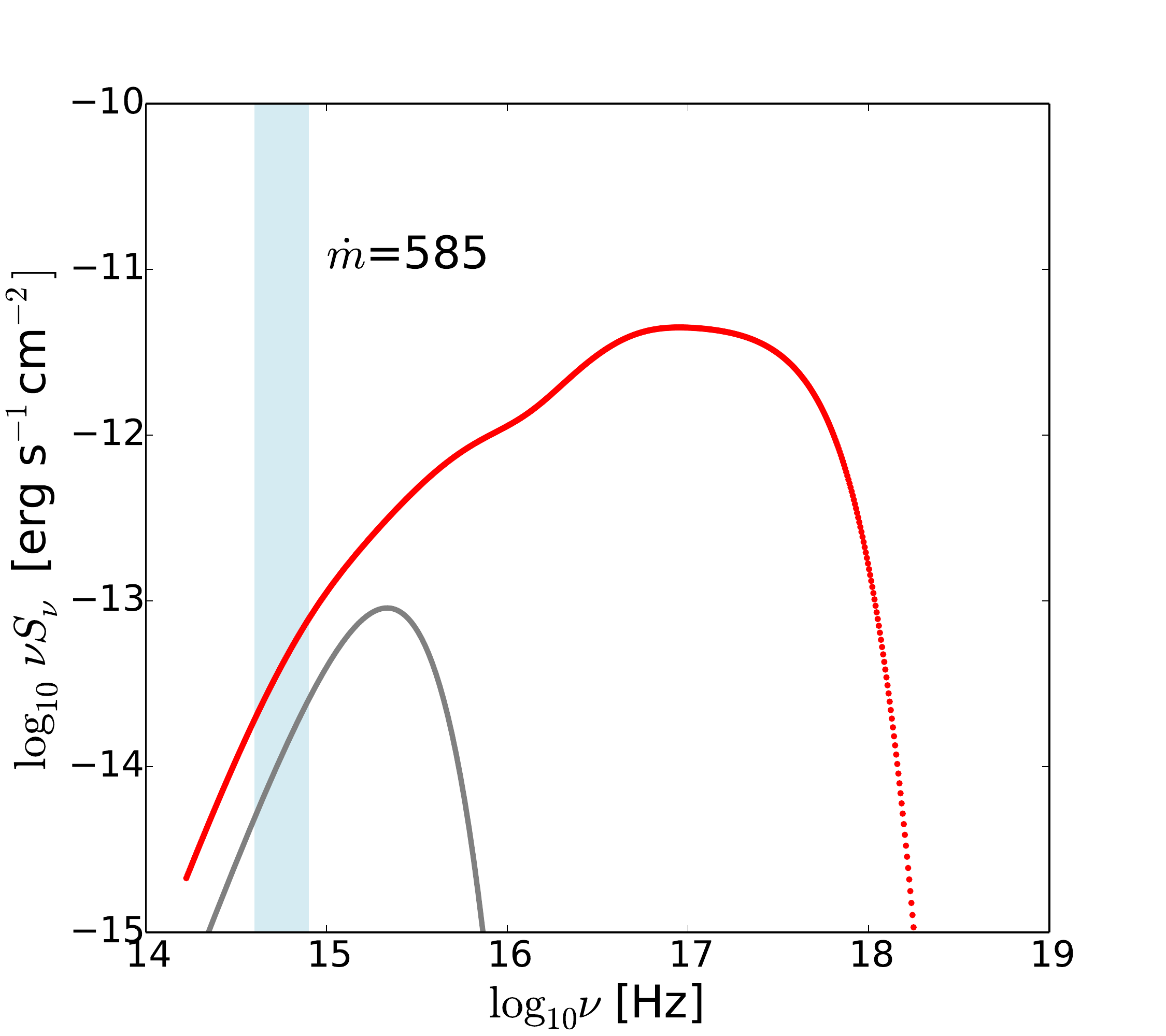}
\task
\includegraphics[width = 0.3\textwidth]{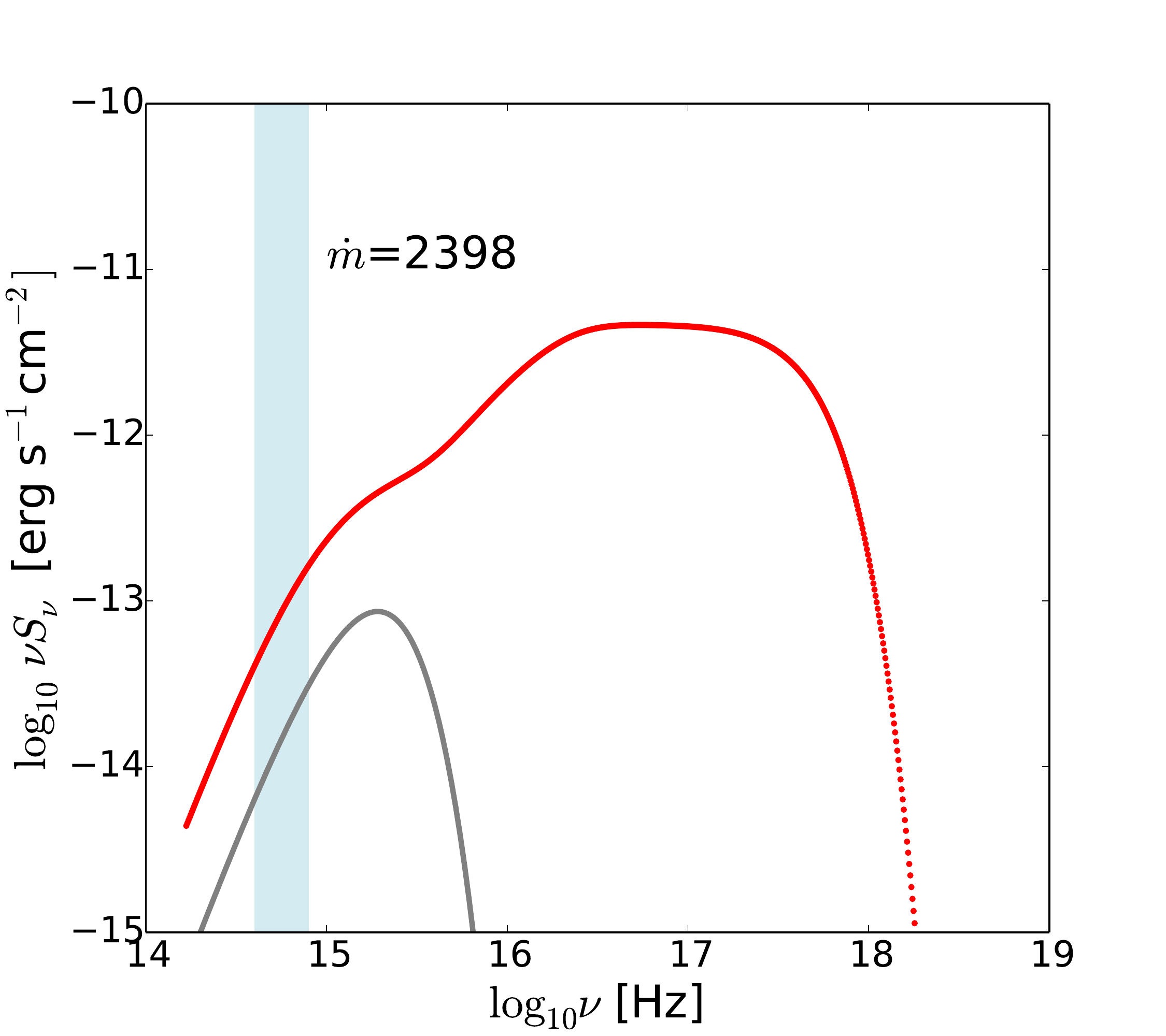}
\task
\includegraphics[width = 0.3\textwidth]{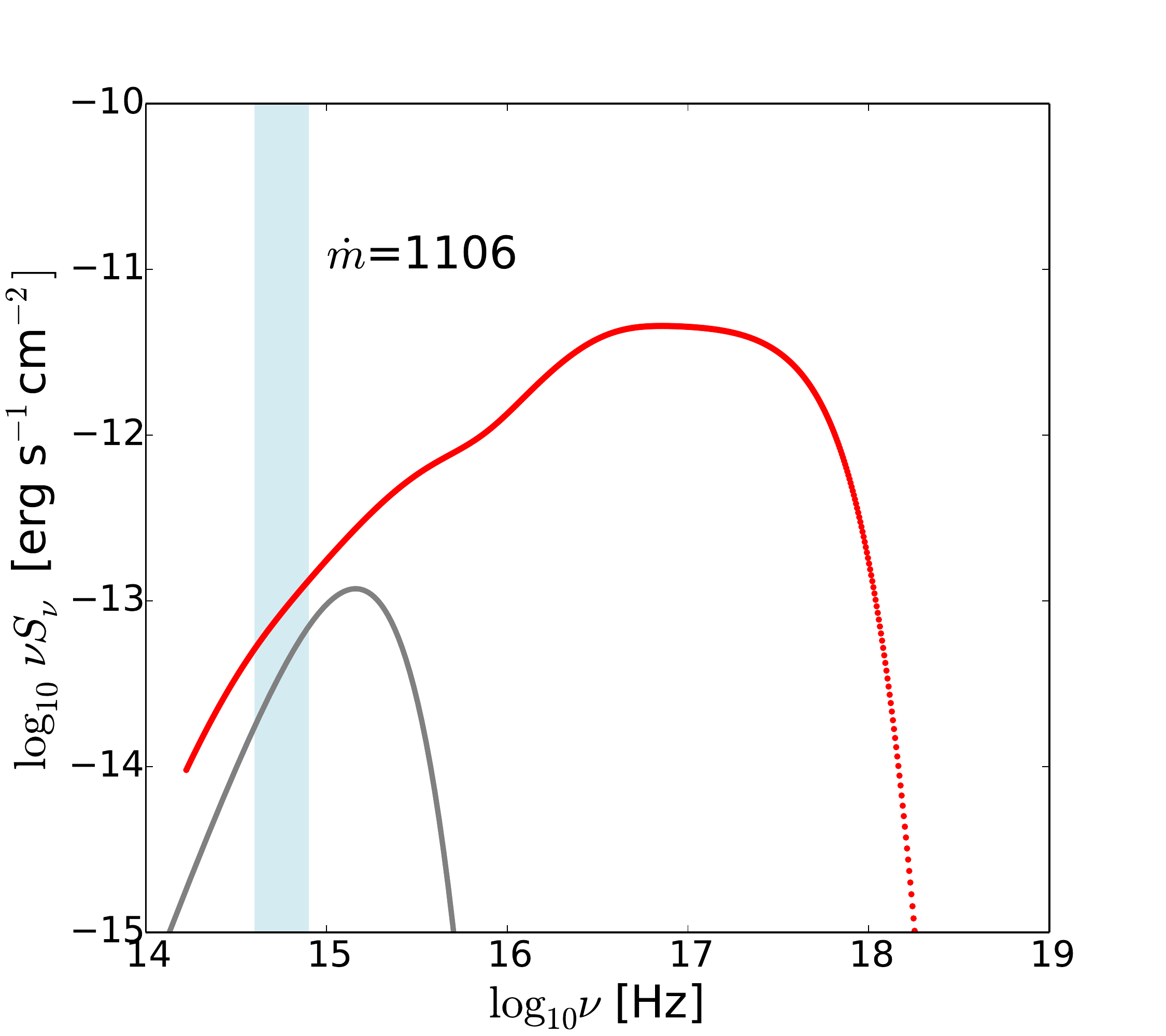}
\end{tasks}
\caption{Same as Figure~\ref{fig:spettro1020*} for a $25 M_{\odot}$ donor and a $100 M_{\odot}$ BH. At phases $a$ and $c$ no irradiation occurs. The mass of the donor star at each time is: $M_{donor,a} \simeq 13.06 M_{\odot}$, $M_{donor,b} \simeq 11.46 M_{\odot}$ and $M_{donor,c} \simeq 7.30 M_{\odot}$.}\label{fig:spettro25100*}
\end{figure*}

In Fig.~\ref{fig:diagrammi*} we show the evolutionary tracks on the color-magnitude diagram calculated in this paper, compared to those reported in PZ (case AB mass transfer). In the lower part of each track the donor is on the Main Sequence (MS). On the upper part the donor has left the MS and moves towards the Giant Branch. During MS $\dot{m}$ is sub-Eddington or mildly super-Eddington, while afterwards accretion becomes highly super-Eddington. As shown in Fig.~\ref{fig:diagrammi*}, during MS our tracks almost overlap those obtained by PZ. The residual difference depends mainly on the different method used for calculating the donor magnitude (see Sec.~\ref{ssec:resphot}) and on the fact that $\dot{m}$ is mildly super-Eddington.

%
When the donor evolves off the MS, the effects induced by the new structure of the accretion flow (super-Eddington accretion with outflow) become relevant and the tracks differ significantly from those of PZ. The evolutionary tracks are brigther and bluer than those computed assuming Eddington-limited accretion.
These features depend on the combination of two effects. First, owing to the larger $\dot{m}$, the flux emitted from the outer standard accretion disc grows significantly. Then, the temperature in the disc and the irradiating flux become progressively higher than in PZ, despite the fact that the extension of the irradiation region is reduced. Second, even if the X-ray-UV emitting region of the disc is hidden below the optically thick outflow, the photosphere of the latter can have a temperature sufficiently high to effectively irradiate the outer regions. These effects increase both the emitted luminosity and irradiation making the system appearing brigther and bluer.

As the system evolves, the orbital separation increases, the accretion disc becomes bigger and the disc emission comparatively redder. In addition, at variance with PZ, in our model we allow the mass ratio to vary during the evolution of the binary system. As all our evolved donors lose a considerable amount of mass, the separation and the accretion disc are bigger than in PZ.
We note that the track of the system with a $25 M_{\odot}$ donor and a $20 M_{\odot}$BH was interrupted when $r_{ph,out}>r_{out}$, as the outflow starts to engulf the binary because of the very high mass transfer and outflow rates, and the model is no longer self-consistent. While this does not rule out the possibility that these systems exist and may reproduce some of the observed ULXs, they are likely to evolve in a significantly different way and treating them is beyond the goals of the present investigation.

In general, systems with a $20 M_\odot$ BH and $12-20 M_\odot$ donors accreting supercritically are characterized by rather blue colors (F450W -- F555W $\simeq - 0.2: + 0.1$) and high luminosity ($M_V \simeq - 4 : - 6.5$) when the donor has evolved off the main sequence. Systems with a massive ($100 M_\odot$) BH accreting supercritically from evolved donors of similar mass have comparable colours but can reach $M_V \simeq - 8$.

Points $a$, $b$ and $c$ on the colour-magnitude diagram in Figure~\ref{fig:diagrammi*} mark the post-MS evolutionary phases at which optical-through-X-ray spectra are computed. They are shown in Figures~\ref{fig:spettro1020*} -~\ref{fig:spettro25100*} and refer to phases with high mass transfer rate. The X-ray spectrum includes the contribution of the disc and the outflow, assuming that they emit locally as blackbodies at different temperatures. A more realistic description of the inner disc spectrum, which involves modelling additional physical processes, will be presented in a forthcoming paper. The bump at optical wavelengths that forms in standard accretion discs because of self-irradiation (e.g. \citealt{1993PASJ...45..443S}) is suppressed as irradiation is comparatively less important than the intrinsic disc emission at very high $\dot{m}$.
The flux emitted from the accretion disc overcomes that from the star in almost all cases.

\subsection{Comparison with the HST photometry of NGC 1313 X-2 and NGC 4559 X-7}
\label{ssec:cmd_outflow_counterparts}


A detailed comparison of the computed evolutionary tracks and optical-through-X-ray spectra of our models with a sample of ULXs is ongoing. Here we show a preliminary application to NGC 1313 X-2 and NGC 4559 X-7 using only optical data. We compare the position on the CMD of the optical counterparts of NGC1313 X-2 and  NGC4559 X-7 with the evolutionary tracks of our systems. We account for ULXs variability taking the mean values of the magnitudes collected in different epochs. Absolute magnitudes are calculated for the distances reported in the literature and adopted for the parent population studies (see below).

\subsubsection*{NGC 1313 X-2}

\begin{figure*}
	\includegraphics[width = \textwidth]{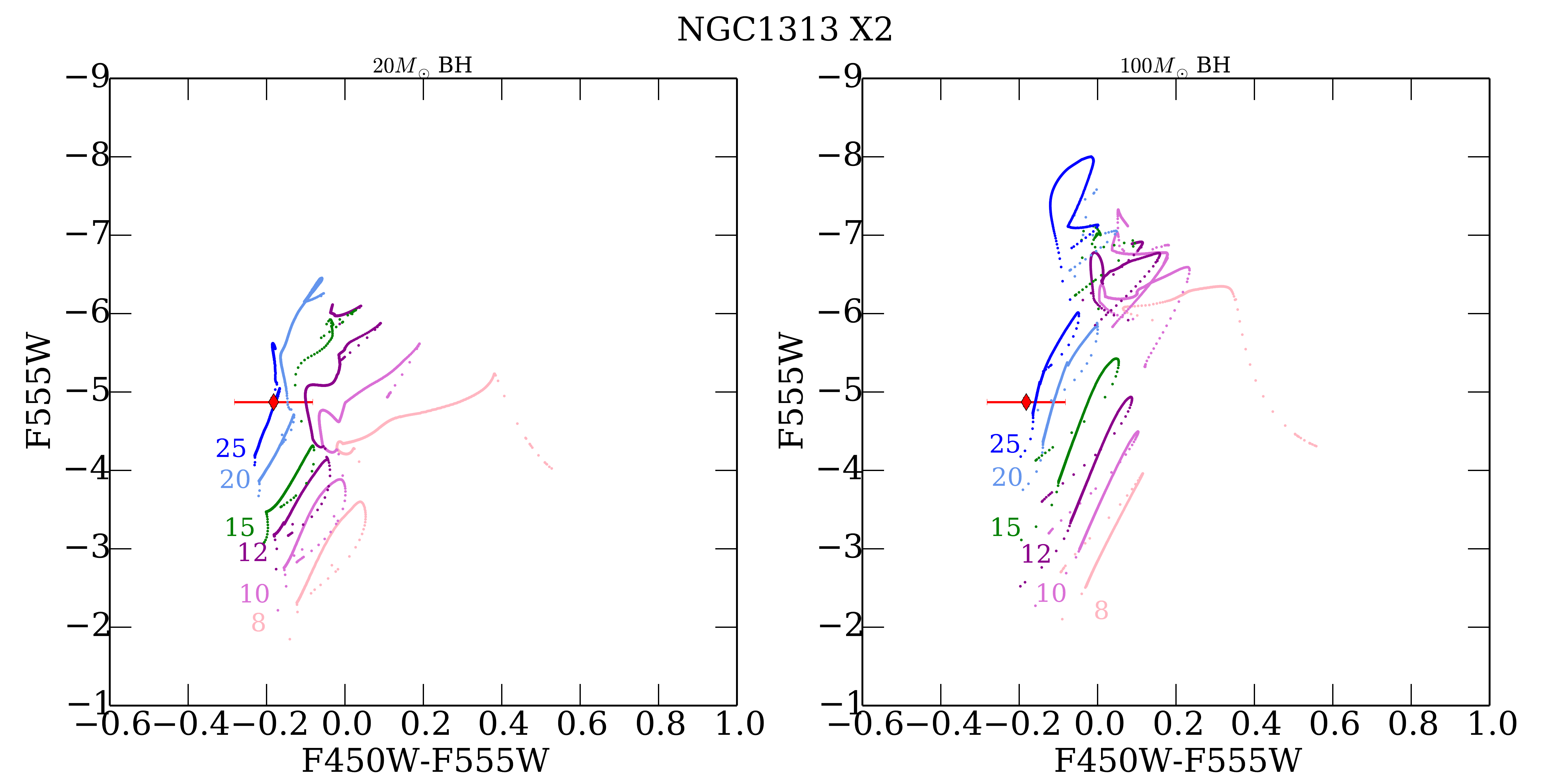}
	\caption{Evolution on the colour-magnitude diagram of ULX binary systems with a BH mass of $20 M_{\odot}$ ({\it left}) and $100 M_{\odot}$ ({\it right}) for donors with different initial main sequence masses (from $8 M_{\odot}$ up to $25 M_{\odot}$). Magnitudes are in the {\it HST} photometric system. The point represents the mean value of all the measured magnitudes of NGC 1313 X-2 (see text for details).  We assume a distance of 4.07 Mpc.}
\label{fig:ngc1313x2}
\end{figure*}

\begin{figure*}
	\includegraphics[width = \textwidth]{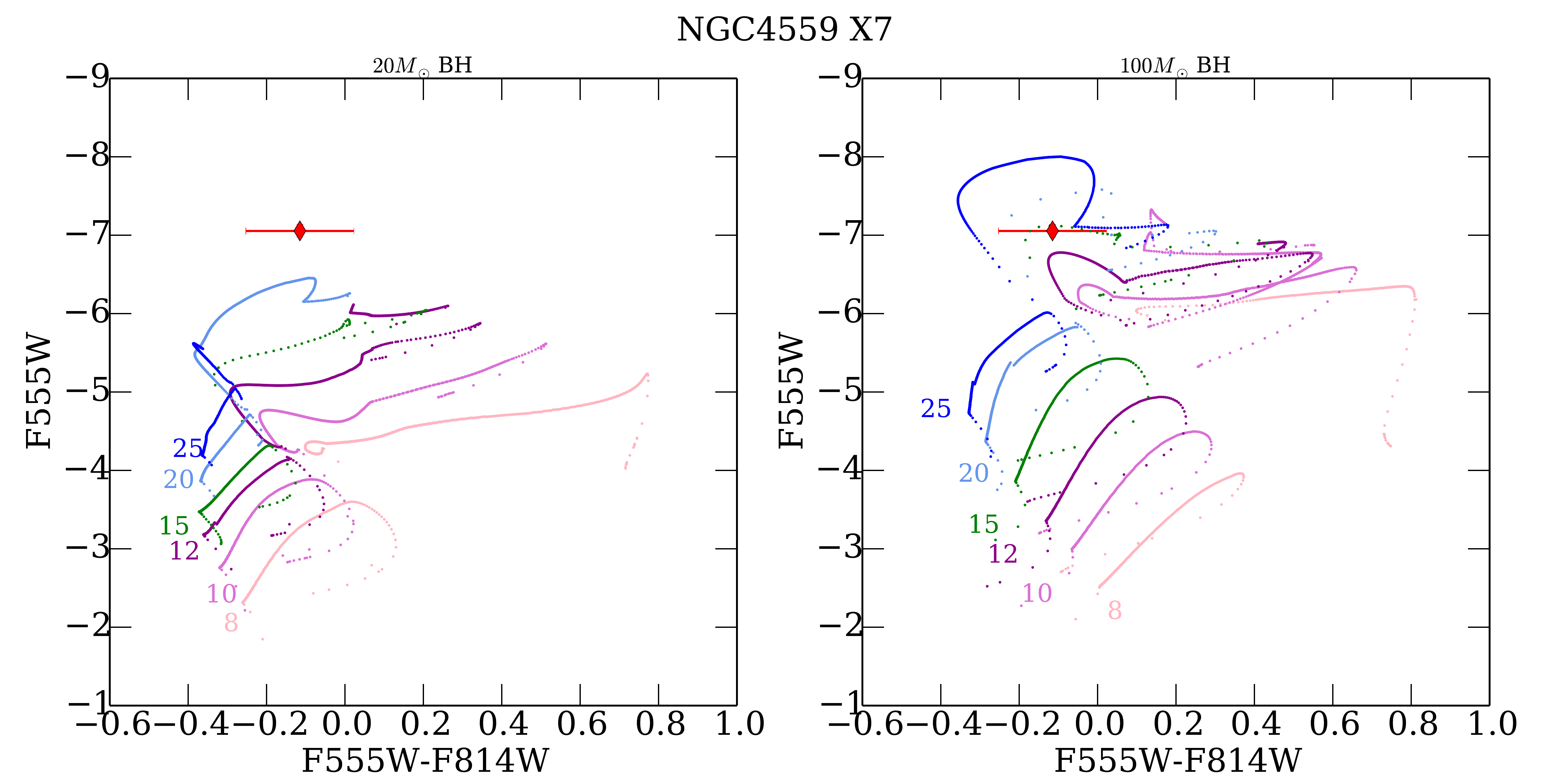}\caption{Same as Fig.~\ref{fig:ngc1313x2} for NGC4559 X-7. We assume a distance of 10 Mpc.}\label{fig:ngc4559x7}
\end{figure*}

NGC 1313 X-2 (hereafter X-2) is one of the best known and studied ULXs, hosted in the barred spiral galaxy NGC 1313. In the following, we assume a distance of $4.07 \pm 0.22$ Mpc (\citealt{2008A&A...486..151G}) and an optical extinction $E(B-V)=0.13$ (\citealt{2011ApJ...737...81T}). The observed X-ray luminosity of NGC 1313 X-2 is in the range $10^{39}$--$3 \times 10^{40}$ erg s$^{-1}$ (\citealt{2007ApJ...658..999M}). The environment of this ULX has been analyzed in detail (\citealt{2004ApJ...603..523Z, 2007ApJ...658..999M, 2008A&A...486..151G}). 
The source is found to reside in an OB association aged $\sim 20 \pm 5$ Myrs \citep{2008A&A...486..151G}. Intensive {\it HST} monitoring of its optical counterpart in 2009 led to the determination of the orbital period of the system, which turns out to be $P = 6.12 \pm 0.16$ days in case of significant X-ray irradiation of the donor or $P = 12.24 \pm 0.16$ if X-ray irradiation is negligible (\citealt{2009ApJ...690L..39L}, \citealt{2012MNRAS.419.1331Z}). As the source appears to show both regular and stocastic variability at the level of $\sim 0.1$ mag, in the following we will consider the average of all measurements of X-2 taken during the 2009 monitoring campaign (F555W = $-4.88 \pm 0.06$, F450W-F555W = $-0.18 \pm 0.1$; \citealt{2011ApJ...737...81T})\footnote{The error on the colour is the quadratic sum of the standard deviations $\sigma = (\sum_i (m_i- {\bar m})^2/(n-1) )^{1/2}$ of the single magnitudes, where we take $\sigma$ (not the error of the mean) as representing the dispersion of the measurements induced by the intrinsic source variability.}. This value then represents a reasonable smoothed average of the observed orbital and X-ray irradiation-induced variability.

In Fig.~\ref{fig:ngc1313x2} we compare the position on the colour-magnitude diagram of the optical counterpart of X-2 with the evolutionary tracks of ULX binaries accreting above Eddington. The point representing X-2 intersects: (i) the tracks of a $100 M_{\odot}$ BH with donors of 20 and $25 M_{\odot}$ in the MS phase, (ii) the tracks of a stellar-mass BH with donors of 12, 15, 20 and $25 M_{\odot}$ during the terminal age MS or H-shell burning phase, when accretion is super-critical. We rule out the first possibility because the duration of the entire tracks that intersect the photometric point is less than 8--12 Myrs, not in agreement with the age of the parent stellar population ($\sim 20 \pm 5$ Myrs). Concerning the stellar-mass BH case (ii), we exclude donors more massive than $20 M_{\odot}$ for similar reasons. On the other hand, case (ii) with donors having initial masses in the range $12-15 M_{\odot}$ is in agreement with the age of the parent stellar population. For these systems the bolometric luminosity ($\sim 10^{40}$ erg s$^{-1}$) is consistent with the observed range of values. It is important to note that, at the time of intersection, accretion is highly super-critical and the optical emission is dominated by the outer standard portion of the disc. Nonetheless, the contribution of the star to the optical flux is 10\%, sufficient to produce a $\sim 0.1$ mag modulation on the optical light curve if irradiation is important. The orbital period P is between 4.5 and 5 days, in reasonable agreement with the reported photometric periodicity. In this case, the observed $\sim 6$ days modulation would represent a single orbital period and the maximum of the light curve would correspond to superior conjunction, when the largest fraction of the irradiated donor surface is visible. Any evidence of a $\sim 12$ days asymmetric modulation would rule out this scenario.

\subsubsection*{NGC 4559 X-7}

NGC 4559 X-7 (hereafter X-7) is a ULX located in the outerskirts of the spiral galaxy NGC 4559. We assume a Galactic extinction $E(B-V) = 0.018$ \citep{2011ApJ...737...81T} and a distance of 10 Mpc \citep{2005MNRAS.356...12S}. The ULX is embedded in a young star-forming complex with age $\leq 30$ Myrs and stars in the \textit{Chandra} error circle are Red or Blue Supergiants with age $\sim 20$ Myrs. The average magnitude and colours of the optical counterpart of X-7 are F555W = $-7.05\pm0.08$ and F555W-F814W = $-0.11\pm0.13$ (from two measurements; \citealt{2011ApJ...737...81T}), while its bolometric luminosity is $\sim 5 \times 10^{40}$ erg s$^{-1}$ \citep{2005MNRAS.356...12S}.

In Fig.~\ref{fig:ngc4559x7} we compare the position on the colour-magnitude diagram of the optical counterpart of X-7 with the evolutionary tracks of ULX binaries accreting above Eddington. We rule out that X-7 is powered by accretion onto a stellar-mass BH of 20 $M_{\odot}$ because, even for super-Eddington accretion from a $20-25 M_\odot$ donor in the H-shell burning phase, the magnitude is too low to reproduce the observed photometry. Conversely, the point representing X-7 intersects the evolutionary tracks of a 100 $M_{\odot}$ BH accreting above Eddington from a $15-25 M_{\odot}$ evolved donor. Assuming a characteristic age of $\sim 20$ Myrs, a $\sim 15 M_\odot$ donor is more plausible. Tracks computed for a more extended grid of BH masses show agreement for masses in the range $70-100 M_{\odot}$ (Ambrosi, Zampieri et al. in prep.). For all these systems the bolometric luminosity ($\sim 4 \times 10^{40}$ erg s$^{-1}$) is in fair agreement with the observed value.\\


Finally, we note that, irrespective of the BH mass, both the counterparts of NGC 1313 X-2 and NGC 4559 X-7 appear to be associated with evolved donors. Despite the post-MS phase of a massive star not being its longest evolutionary phase, it is certainly the only one with significant super-Eddington rates. Limiting the comparison to this phase, the position on the post-MS tracks of NGC 1313 X-2 (where agreement with the observed photometry is found) falls on a rather short-lived phase of the system (when the orbital separation is rapidly increasing), while it falls along a comparatively slower evolutionary phase for NGC 4559 X-7. On this ground alone, the probability of finding NGC 4559 X-7 at the observed position on the CMD seems somewhat higher than that of NGC 1313 X-2.

\section{Discussion and conclusions}


In this work we modelled the optical emission of ULXs accreting super-Eddington. At first we considered a bimodal accretion disc formed by an inner advection-dominated slim disc and an outer standard disc. Then, we included the effects of an outflow produced by radiation pressure. As noted above, the case of a pure slim disc is useful to understand the effects of supercritical accretion on the emission properties of the disc and on self-irradiation. For $\dot{m} > 10^{3}$ the outflow can completely cover the accretion disc and the evolution can no longer be followed with the present model.

Super-critical accretion has considerable effects on the optical emission. The disc self-irradiation is very different from that produced in the standard case. The innermost regions do not radiate the outer ones. Two competing effects influence self-irradiation. On one side, the irradiating flux grows with $\dot{m}$. On the other side, it decreases because the size of the irradiating region diminishes. Irradiation is considerably stronger than that produced by an Eddington-limited disc, but at very high $\dot m$ it is progressively less important than the intrinsic disc and donor emission.

During MS, the evolutionary tracks on the CMD obtained with our model almost overlap with those of \cite{2008MNRAS.386..543P}. The residual difference depends on the treatment of photometry and to the fact that the accretion rate can be mildly super-Eddington. On the other hand, the post MS evolution is markedly different and is characterized by two phases. Initially, when the accretion disc is not very extended, the luminosity increases with $\dot m$ and the system becomes bluer ($B-V = -0.3-0.1$). As the orbital seperation increases, the accretion disc becomes bigger, the emission becomes progressively redder and the system moves to the right on the CMD. At super-Eddington rates, the disc flux typically overcomes that produced by the donor (apart from the case of small black holes and very massive donors) and the 'optical bump' that characterizes standard self-irradiated discs disappears.



We compared the predictions of our model with the {\it HST} observations of NGC 1313 X-2 and NGC 4559 X-7. We constrain the donor and the BH mass of these objects, considering as a further constraint the age of the parent population and the observed bolometric luminosity. The position of NGC 1313 X-2 on the CMD is in agreement with a $20 M_{\odot}$ BH accreting above the Eddington limit from a post-MS donor with initial mass in the range $12-15 M_{\odot}$. The orbital period of the system is $\sim 5$ days, in fair agreement with the observed one. It should be emphasized that the $20 M_{\odot}$ BH and $12-15 M_{\odot}$ donor tracks intersect the observed photometric point of X-2 because accretion is highly super-critical and the optical emission, dominated by the outer standard portion of the disc, is significantly enhanced and blue in comparison with that for Eddington-limited accretion.

We now compare the results of the dedicated analysis of NGC 1313 X-2 presented by \cite{2010MNRAS.403L..69P} (PZ2) for Eddington-limited accretion with those obtained here. 
PZ2 found that the photometric point representing X-2 is well reproduced by a system with a $\sim 50-100 M_\odot$ BH and a MS donor of $\sim 12-15 M_\odot$ or by a system with a $\sim 20 M_\odot$ BH undergoing mass transfer from a H-shell burning donor of $\sim 12-15 M_\odot$. The orbital period of the systems that intersect X-2 on the CMD is $\sim 6$ days in the first case and $\sim 12$ days in the second case. In both cases accretion is Eddington-limited. 
A similar conclusion for the $\sim 20 M_\odot$ BH case was reached by \cite{2014ApJ...794....7M} who calculated the magnitudes of simulated ULX binaries evolved in young star clusters and compared them with the available photometry of several ULXs. The adopted code is that of PZ and PZ2, while the input parameters at the beginning of the Roche-lobe overflow phase (radius, mass, optical luminosity, effective temperature, and age of the donor, mass of the BH, orbital period) were provided by $N$-body simulations, that did not produce matches with systems having BH masses larger than $\sim 25 M_\odot$.

In fact, besides the accretion regime, there are two other aspects to consider when comparing the results of PZ2 with our ones. First, PZ2 considered the {\it HST} ACS observations performed in 2003 (taken from \citealt{2007ApJ...658..999M}), while here we considered a longer dataset, averaging over the variability induced by the orbital motion and X-ray irradiation. Second, the conversion from {\it HST} ACS/WFC F435W and F555W filters to the standard Johnson photometric system leads to some inaccuracy, especially for variable sources (see e. g. \citealt{2005PASP..117.1049S}).
We note that also \cite{2014ApJ...794....7M} used the first 2003 {\it HST} ACS photometric epoch (taken from \citealt{2013ApJS..206...14G}).

Using the two 2003 {\it HST} photometric measurements adopted in PZ2, we find agreement with the tracks of $20-25 M_\odot$ donors during MS for a $100 M_\odot$ BH and the track of a $20 M_\odot$ donor at terminal age MS for a $20 M_\odot$ BH, that do not have the correct age of the parent OB association. This would rule out both Eddington-limited accretion onto a $100 M_\odot$ BH and super-Eddington accretion onto a $20 M_\odot$ BH for X-2. On the other hand, comparison with the average photometry of the 2009 {\it HST} monitoring campaign still rules out the $100 M_\odot$ BH scenario, but allows for super-Eddington accretion onto a $20 M_\odot$ BH from a H-shell burning donor of $\sim 12-15 M_\odot$. Therefore, adopting the 2009 average photometry and performing the comparison with the new tracks in the original {\it HST} photometric system is crucial to rule out the Eddington-limited $100 M_\odot$ BH scenario of PZ2 and to pinpoint the matching $20 M_\odot$ BH super-Eddington system.

Letting aside the details of the photometric comparison, we note that the orbital period of the $20 M_\odot$ BH matching binary system is $\sim 12$ days for Eddington-limited accretion, while it is $\sim 5$ days for super-Eddington accretion. In fact, in order to reach a comparable optical luminosity an Eddington-limited disc has to be more extended, and hence the system has to have a longer orbital period. This would provide a further means to distinguish between Eddington-limited and super-Eddington accretion in this sytem. Unfortunately, present measurements are not sufficiently accurate to discriminate between a $\sim 6$ days irradiation-modulated orbital periodicity and a $\sim 12$ days ellipsoidally-modulated one.

Also the optical counterpart of NGC 4559 X-7 has been previously analysed by PZ. The same caveat concerning the photometric transformation to the Johnson system applies also here, but is less critical. The same {\it HST} measurement of \cite{2005MNRAS.356...12S} is considered. PZ found that X-7 is reproduced by a massive or a stellar-mass BH accreting from a $30-50 M_{\odot}$ donor during the H-shell burning phase. We did not consider donors more massive than $25 M_{\odot}$ for a 20 $M_{\odot}$ BH because, after the MS, the outflow starts to engulf the binary and the model is no longer self-consistent. For less massive donors, no agreement is found because the optical emission is not sufficiently luminous. For a 100 $M_{\odot}$ BH, agreement is found for evolved donors of $15-25 M_{\odot}$, smaller than in the Eddington-limited systems of PZ owing to the fact that super-Eddington accretion makes the tracks more luminous and bluer for a given donor mass.

We note that our result for NGC 1313 X-2 is consistent with the findings of \cite{2013ApJ...778..163B}, who performed a detailed analysis of the \textit{NuSTAR} and \textit{XMM-Newton} spectra of the source using a slim disc plus Comptonization model and found that they are well described by a low mass BH ($M \sim 20-30 M_{\odot}$).
The results for NGC 1313 X-2 and NGC 4559 X-7 are also in agreement with the BH masses estimated by \citealt{2017MNRAS.469L..99F} using the same wind model \citep{2007MNRAS.377.1187P}. From the properties of the outflow measured in the X-rays they constrain the masses and accretion rates of NGC 1313 X-1, NGC 55 ULX and NGC 5408 X-1, finding that the BH mass is likely to be in the range $10-100 M_{\odot}$.

In the future we will explore in detail the properties of all the ULXs with avaiable optical observations. We will compare them with updated evolutionary tracks covering a wider parameter space in terms of BH and donor masses. The comparison will be performed in the same photometric system used for the observations, accounting for the variability of the sources and including the optical-through-X-ray spectral energy distribution. Moreover, the case of accreting NSs will be explored in order to asses whether optical emission can help distinguishing between ULXs hosting BHs and NSs. As detecting pulsation has proven to be very difficult, it would be important to have other methods to discriminate among them.
Modelling accreting NSs is made more difficult by the dynamical and radiative effects of the magnetic fields, the structure and strength of which in pulsar ULXs are unknown. At present we are working at extending our super-Eddington accretion model to the case of a non-magnetized NS, which can be accomodated with relatively little effort within the present framework. Results will be presented in a forthcoming paper.

\section*{Acknowledgements}
We thank the referee Ciro Pinto for the useful comments. 
We acknowledge financial contribution from the grants ASI-INAF I/037/12/0 (project ``Accretion onto stellar and intermediate-mass compact objects with NuSTAR'') and ASI/INAF n. 2017-14-H.O (projects "High-Energy observations of Stellar-mass Compact Objects: from CVs to Ultraluminous X-Ray Sources`` and ''Understanding the x-ray variabLe and Transient Sky (ULTraS)``).

\newpage




\bibliographystyle{plain}




\appendix

\section{Bimodal disc structure without outflow}
\label{appendix:app}
\begin{figure*}
 \includegraphics[width=\columnwidth]{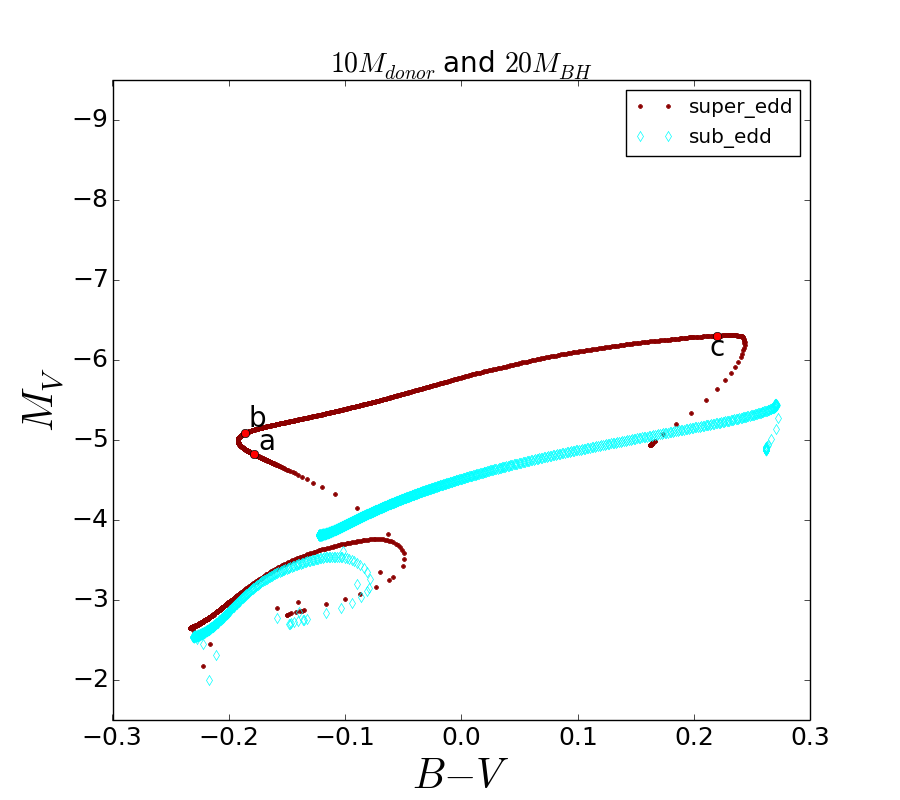}
 \includegraphics[width=\columnwidth]{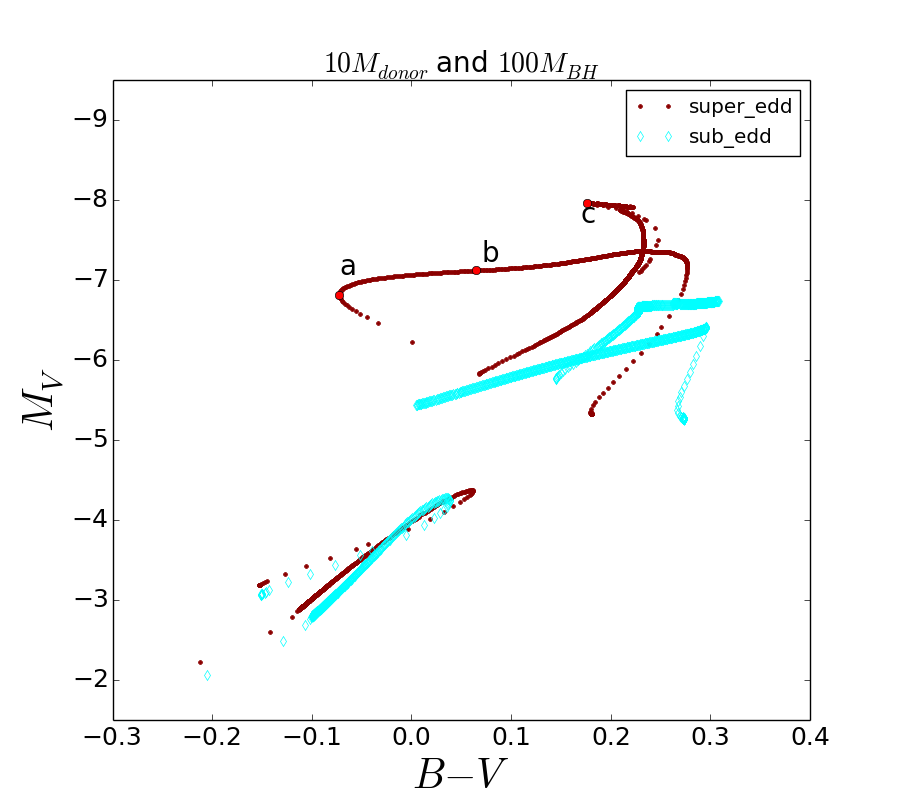}
 \includegraphics[width=\columnwidth]{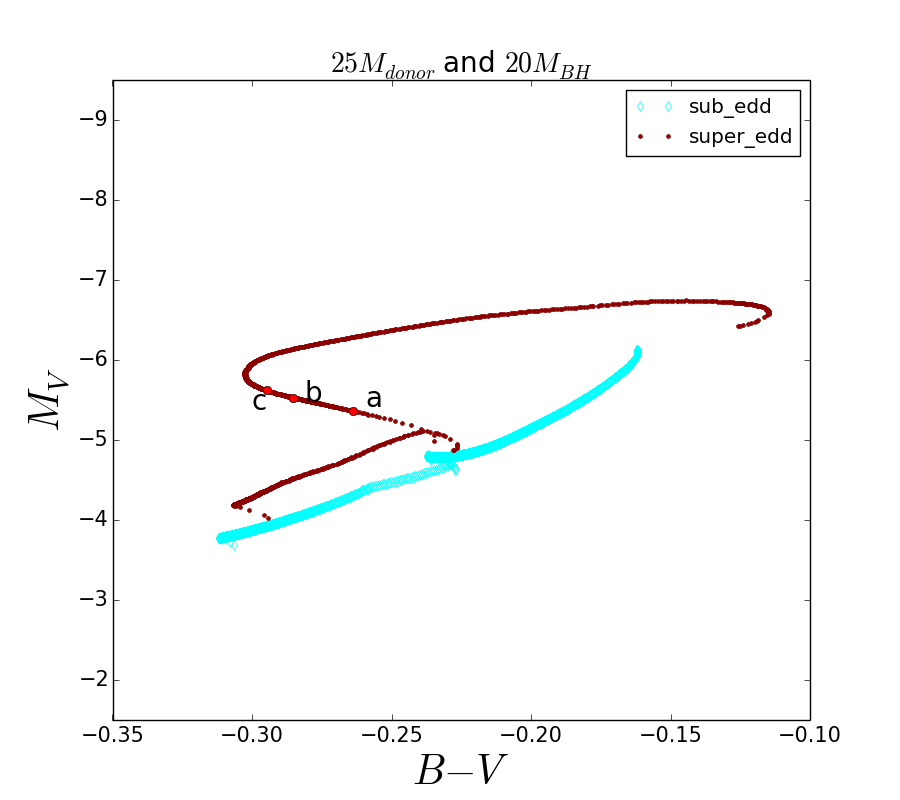}
 \includegraphics[width=\columnwidth]{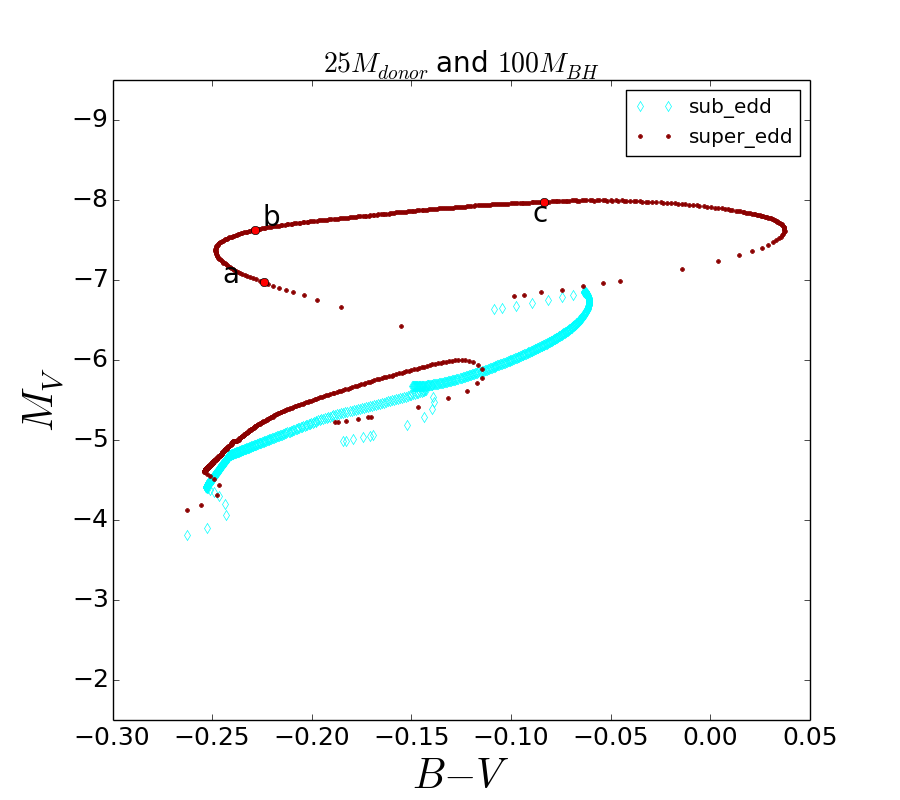}
 \caption{Evolution on the colour-magnitude diagram of a ULX binary system with a donor mass of $10 M_{\odot}$ ({\it upper} panels) or $25 M_{\odot}$ ({\it lower} panels) at zero age main sequence and a BH mass of $20 M_{\odot}$ ({\it left} panels) or $100 M_{\odot}$ ({\it right} panels). The {\it thick} ({\it light grey}, {\it cyan} in the on-line version) line represents the evolution calculated assuming standard sub-Eddington accretion (PZ). The {\it thin} ({\it dark gray}, {\it red} in the on-line version) line represents the new evolution computed for super-Eddington accretion without an outflow. Points $a$, $b$ and $c$ mark the evolutionary phases at which the spectral energy distribution is computed.}
 \label{fig:diagrammi42*}
\end{figure*}

In this section we report for comparison the results obtained for a disc with a bimodal structure without outflow (Sec.~\ref{ssec:advection_no_outflow}). Fig.~\ref{fig:diagrammi42*} shows the evolutionary tracks for systems with donors of 10 and 25 $M_{\odot}$ and BHs of 20 and 100 $M_{\odot}$, compared to the model of PZ.

\begin{figure*}
\bfseries
\begin{tasks}[counter-format = {(tsk[a])},label-offset = {0.8em},label-format = {\bfseries}](3)
\task
\includegraphics[width = 0.3\textwidth]{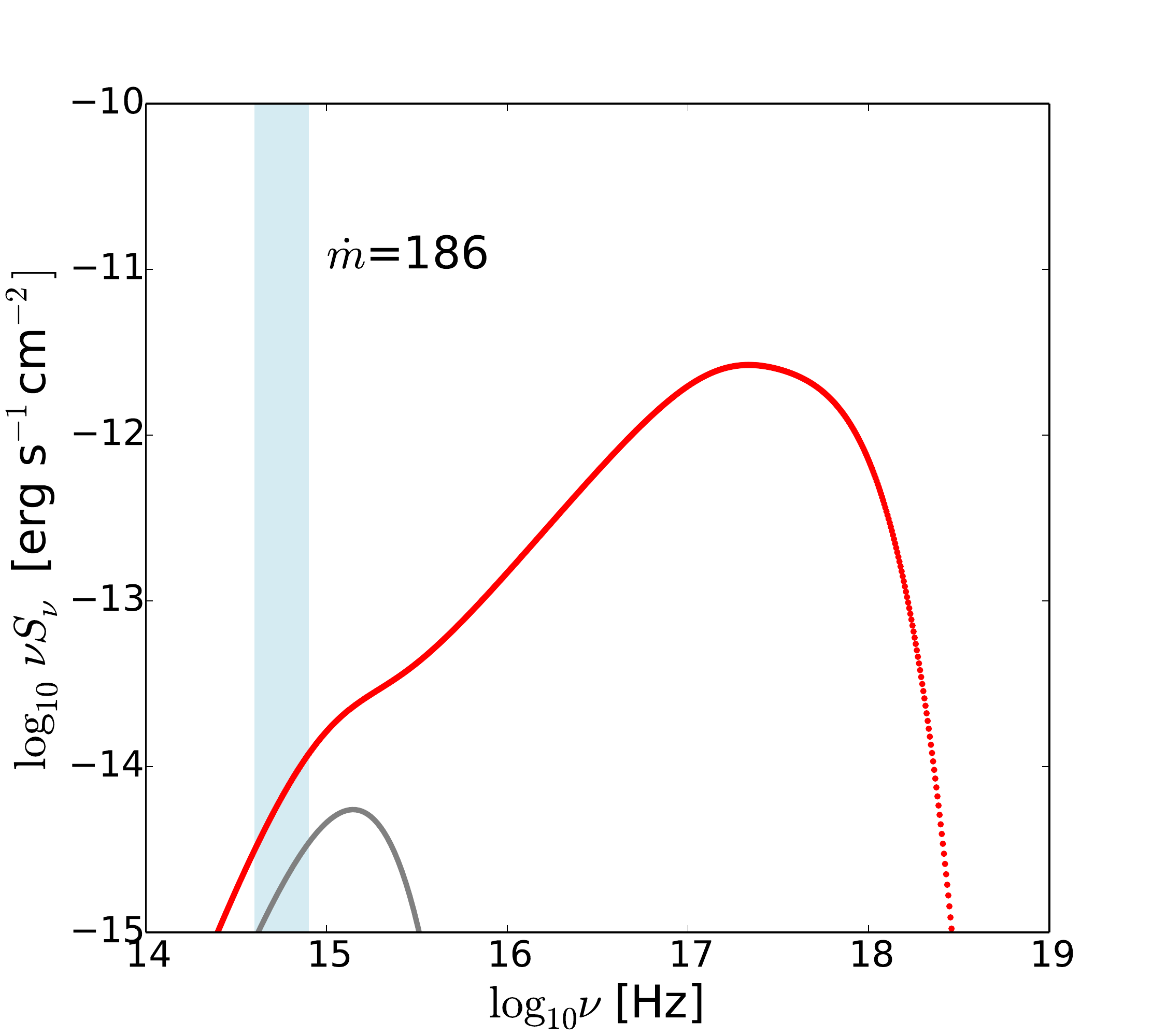}
\task
\includegraphics[width = 0.3\textwidth]{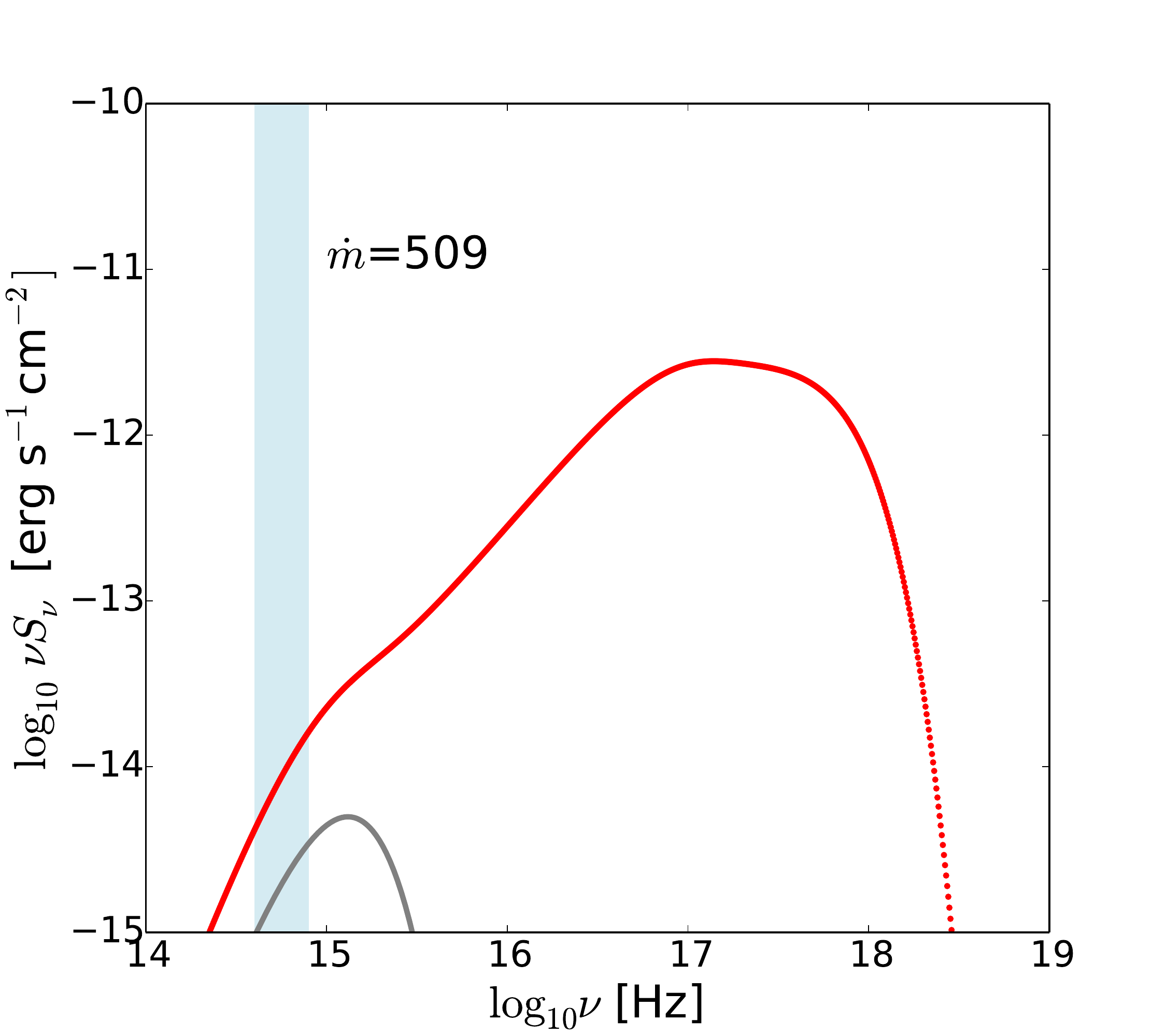}
\task
\includegraphics[width = 0.3\textwidth]{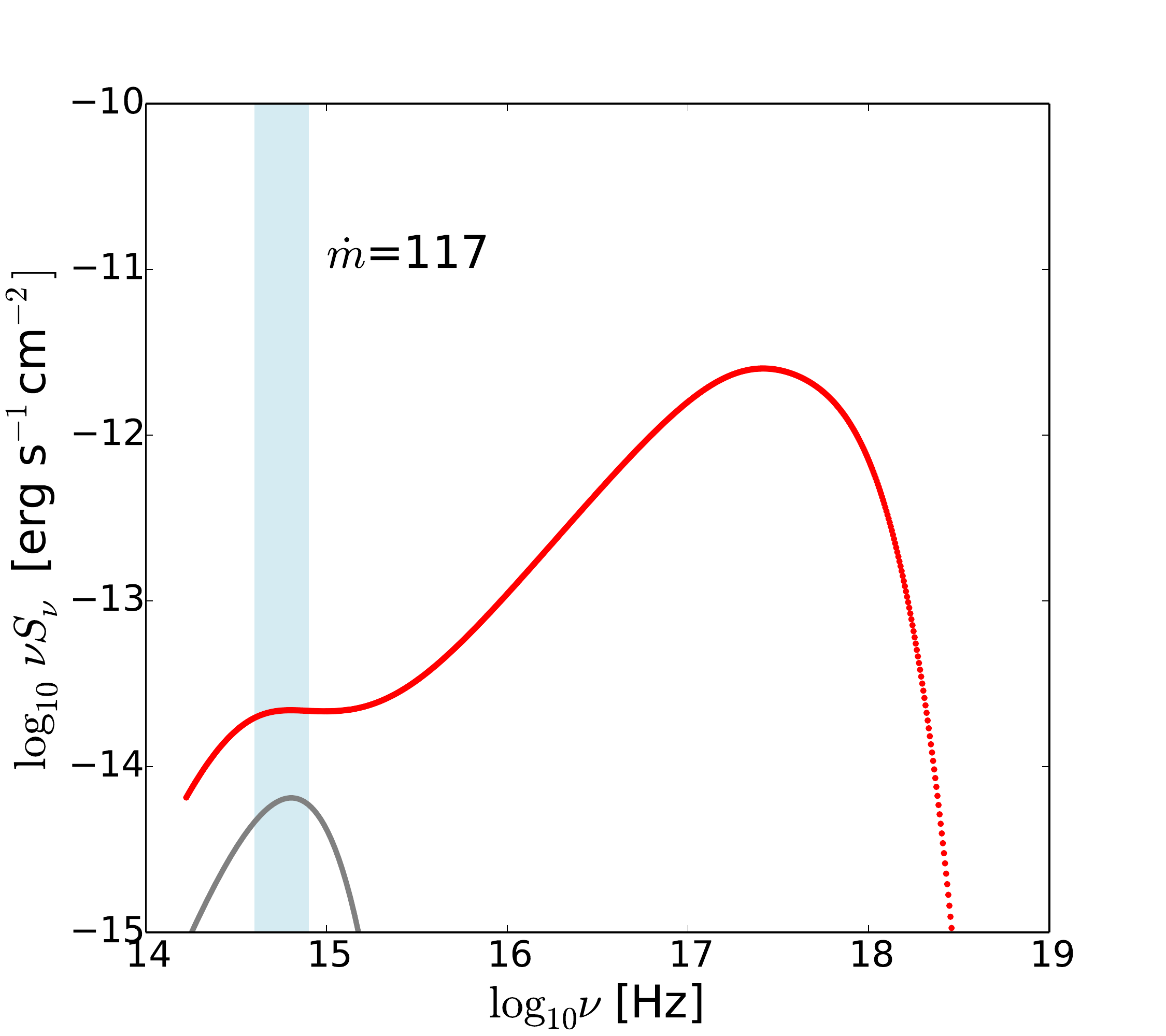}
\end{tasks}
\caption{Optical-through-X-ray spectrum of the system with a $10 M_{\odot}$ donor and a $20 M_{\odot}$ BH at the phases marked with $a$, $b$ and $c$ on the evolutionary tracks for super-Eddington accretion without an outflow. The {\it thick} ({\it red}) line represents the spectrum of the self-irradiated outer disc plus the outflow and the innermost slim disc, while the {\it thin} ({\it gray}) line is the spectrum of the X-ray heated donor. The {\it gray} ({\it light blue}) strip marks the optical band. Irradiation is significant at all phases and causes the observed increase of the optical flux.}
\label{fig:spettro1020_42*}
\end{figure*}

\begin{figure*}
\begin{tasks}[counter-format = {(tsk[a])},label-offset = {0.8em},label-format = {\bfseries}](3)
\task
\includegraphics[width = 0.3\textwidth]{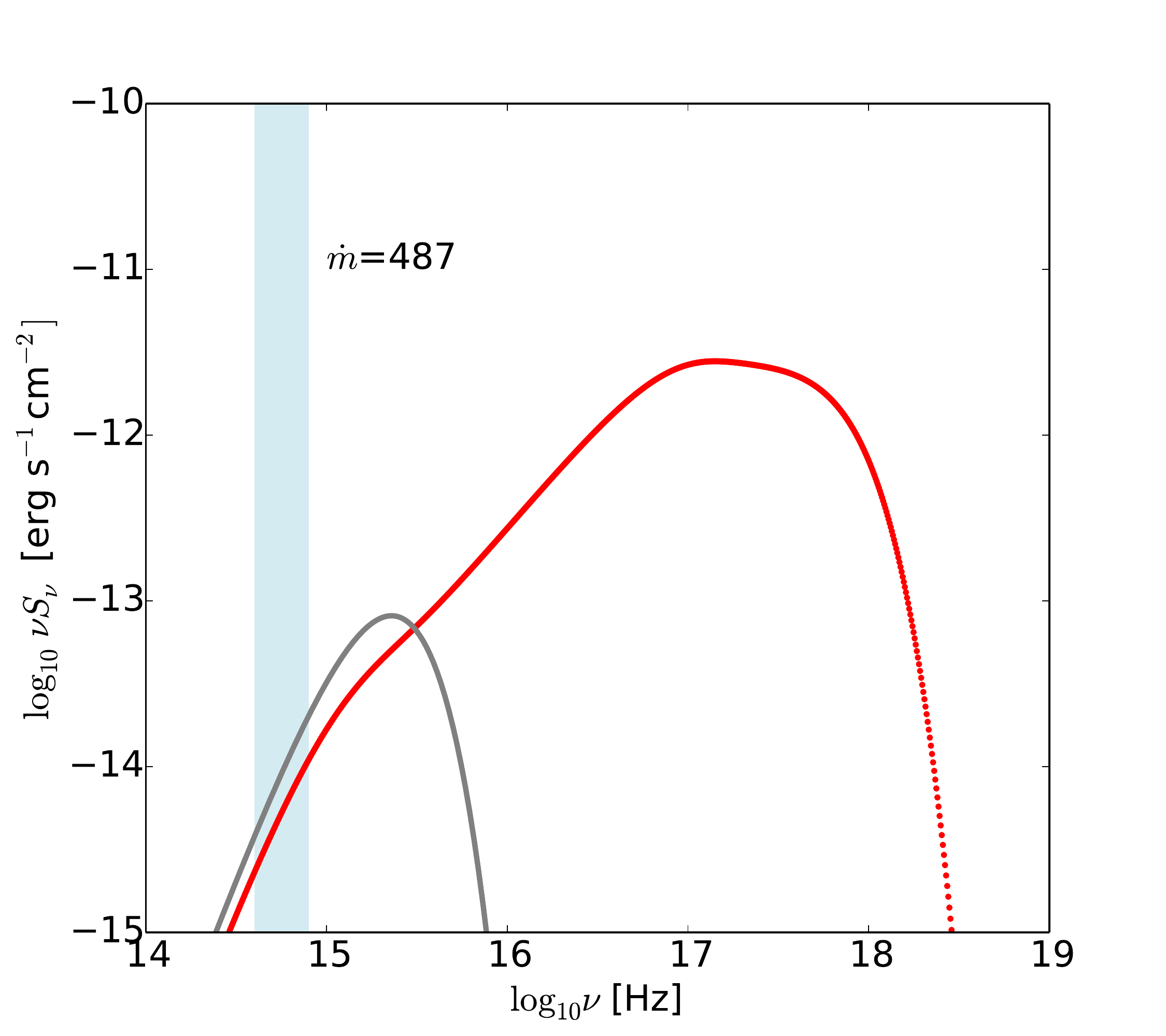}
\task
\includegraphics[width = 0.3\textwidth]{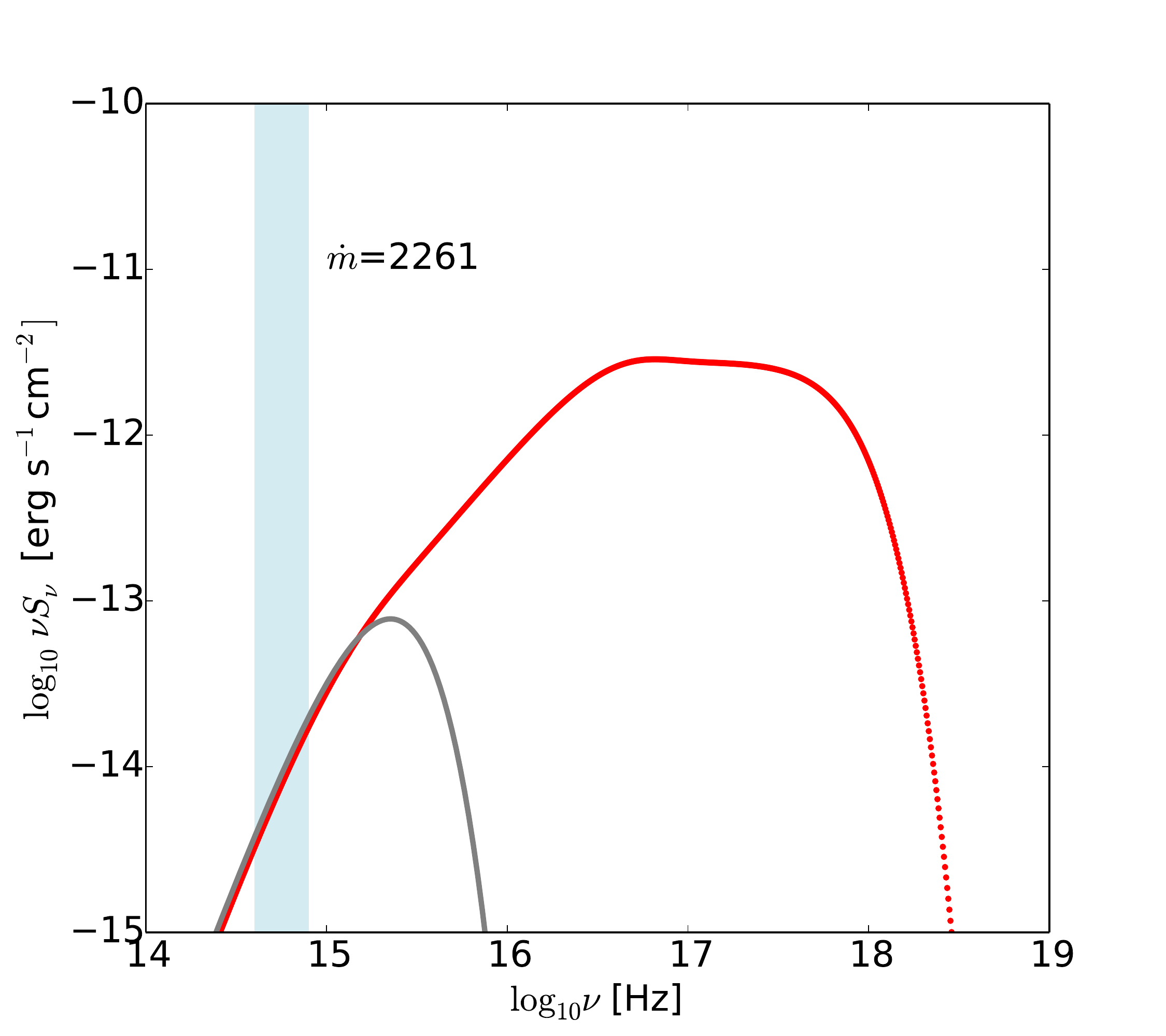}
\task
\includegraphics[width = 0.3\textwidth]{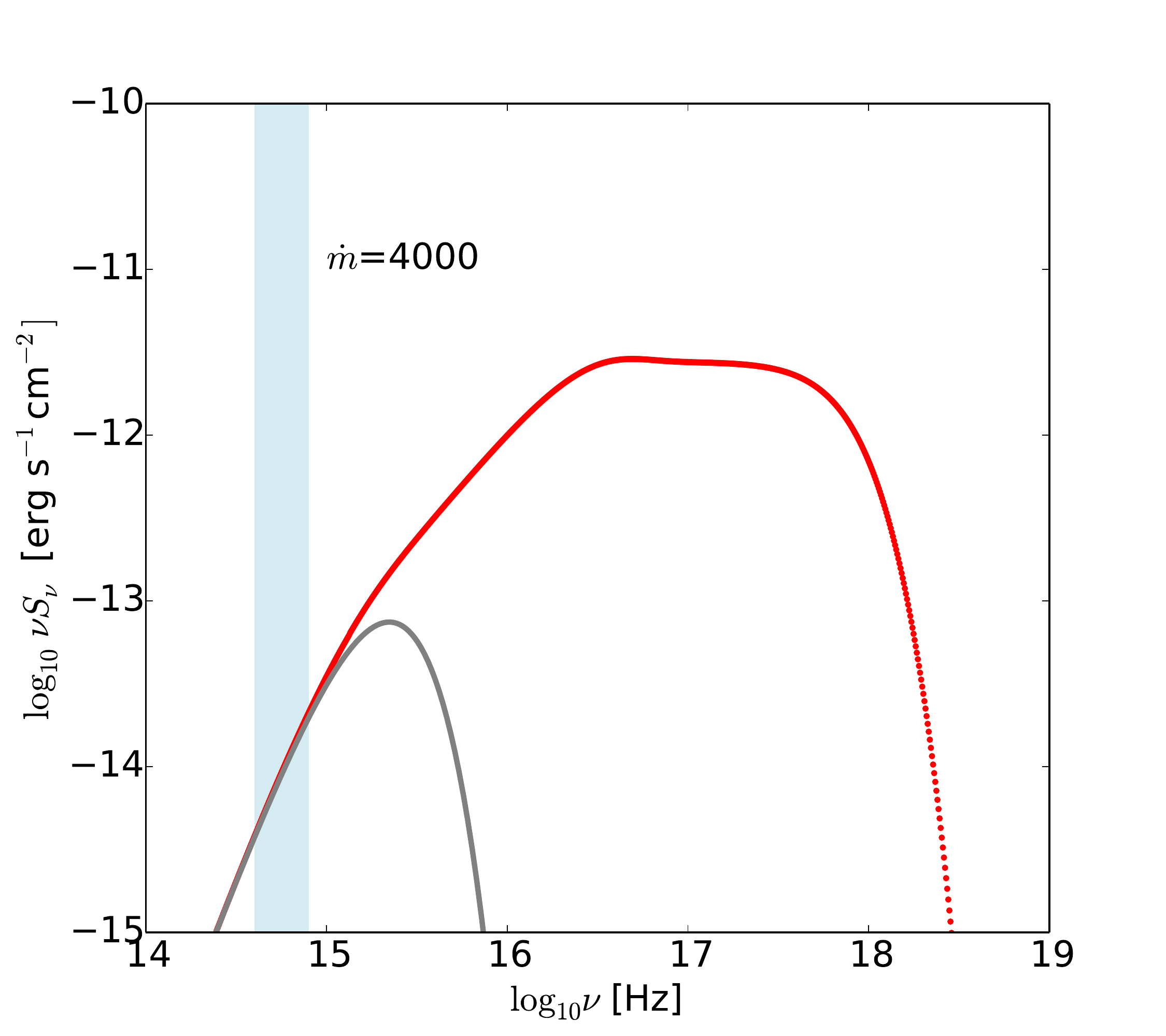}
\end{tasks}
\caption{Same as Figure~\ref{fig:spettro1020_42*} for a $25 M_{\odot}$ donor and a $20 M_{\odot}$ BH. The accretion rate is so high that the optical flux emitted from the disc dominates over that from irradiation.
}
\label{fig:spettro2520_42*}
\end{figure*}

\begin{figure*}
\begin{tasks}[counter-format = {(tsk[a])},label-offset = {0.8em},label-format = {\bfseries}](3)
\task
\includegraphics[width = 0.3\textwidth]{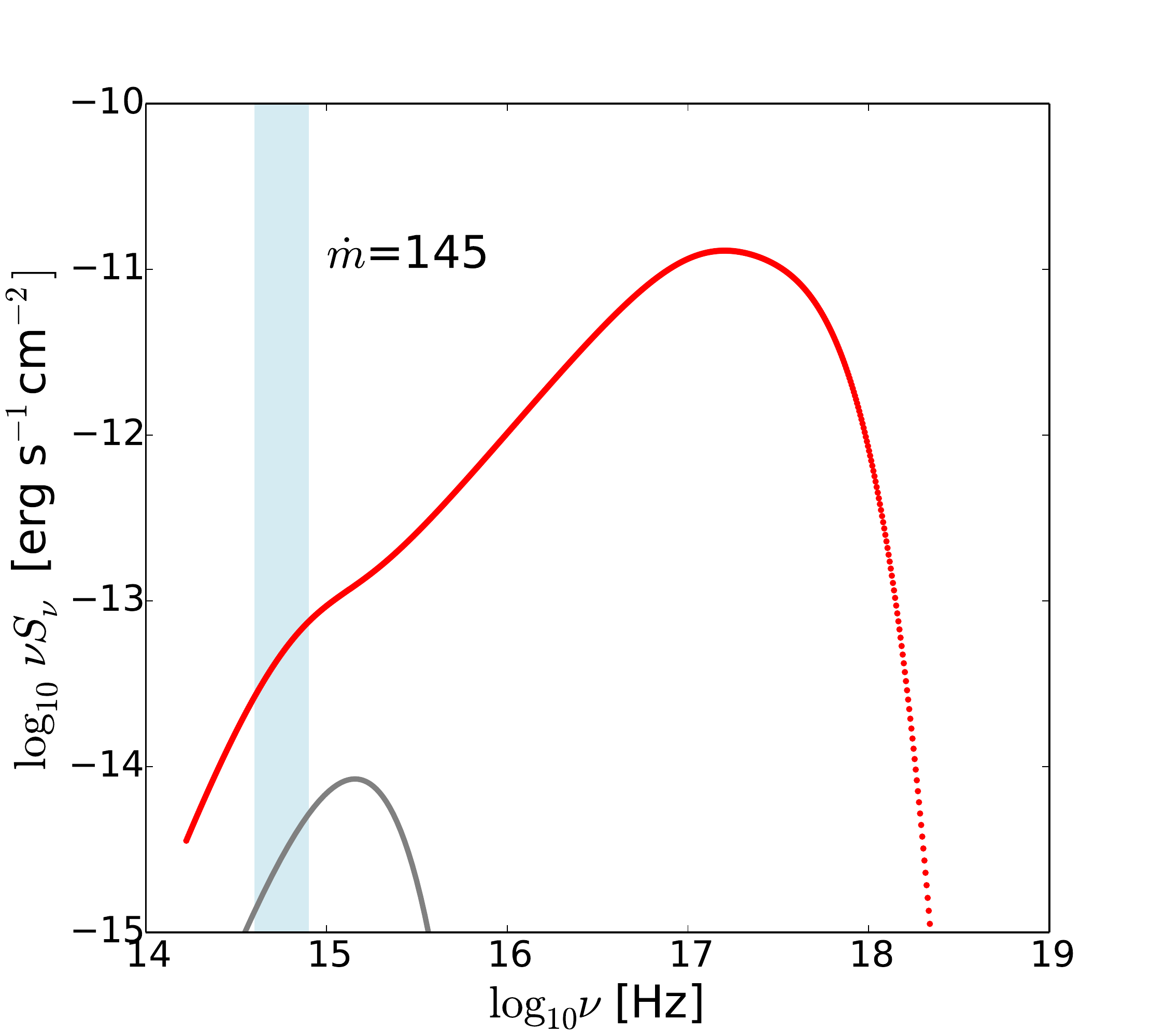}
\task
\includegraphics[width = 0.3\textwidth]{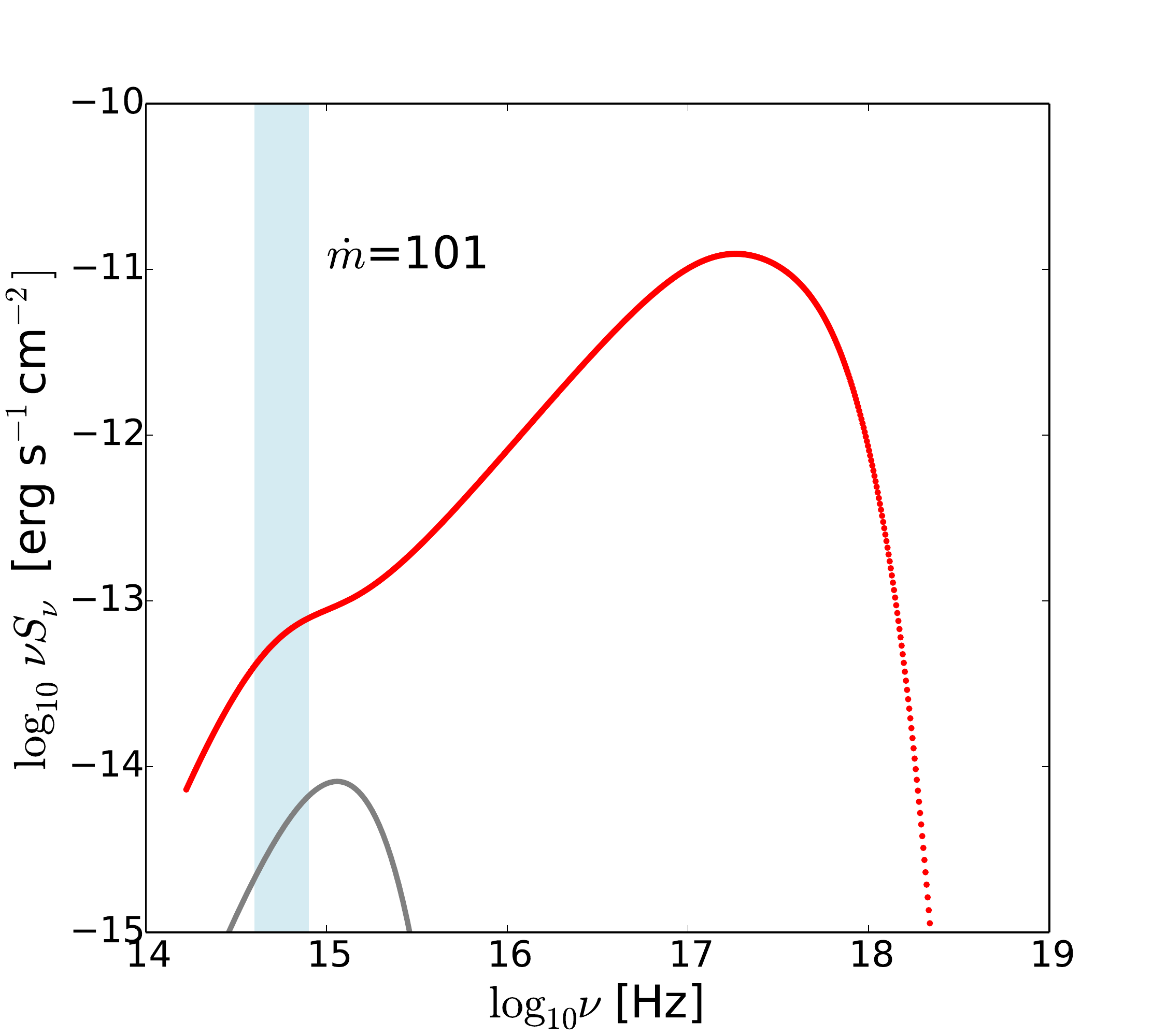}
\task
\includegraphics[width = 0.3\textwidth]{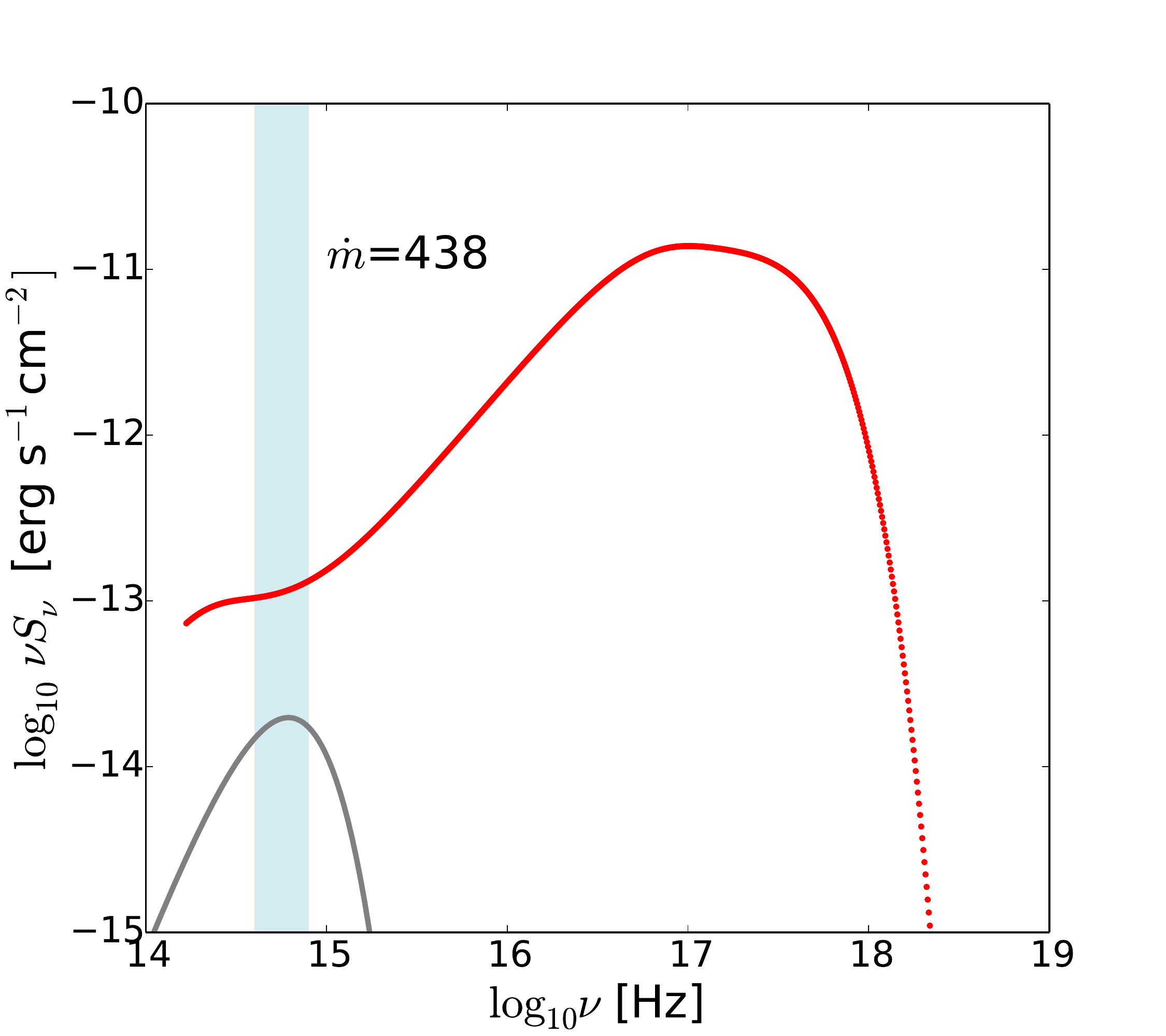}
\end{tasks}
\caption{Same as Figure~\ref{fig:spettro1020_42*} for a $10 M_{\odot}$ donor and a $100 M_{\odot}$ BH. Irradiation is significant at all phases and causes the observed increase of the optical flux.}
\label{fig:spettro10100_42*}
\end{figure*}
\begin{figure*}
\begin{tasks}[counter-format = {(tsk[a])},label-offset = {0.8em},label-format = {\bfseries}](3)
\task
\includegraphics[width = 0.3\textwidth]{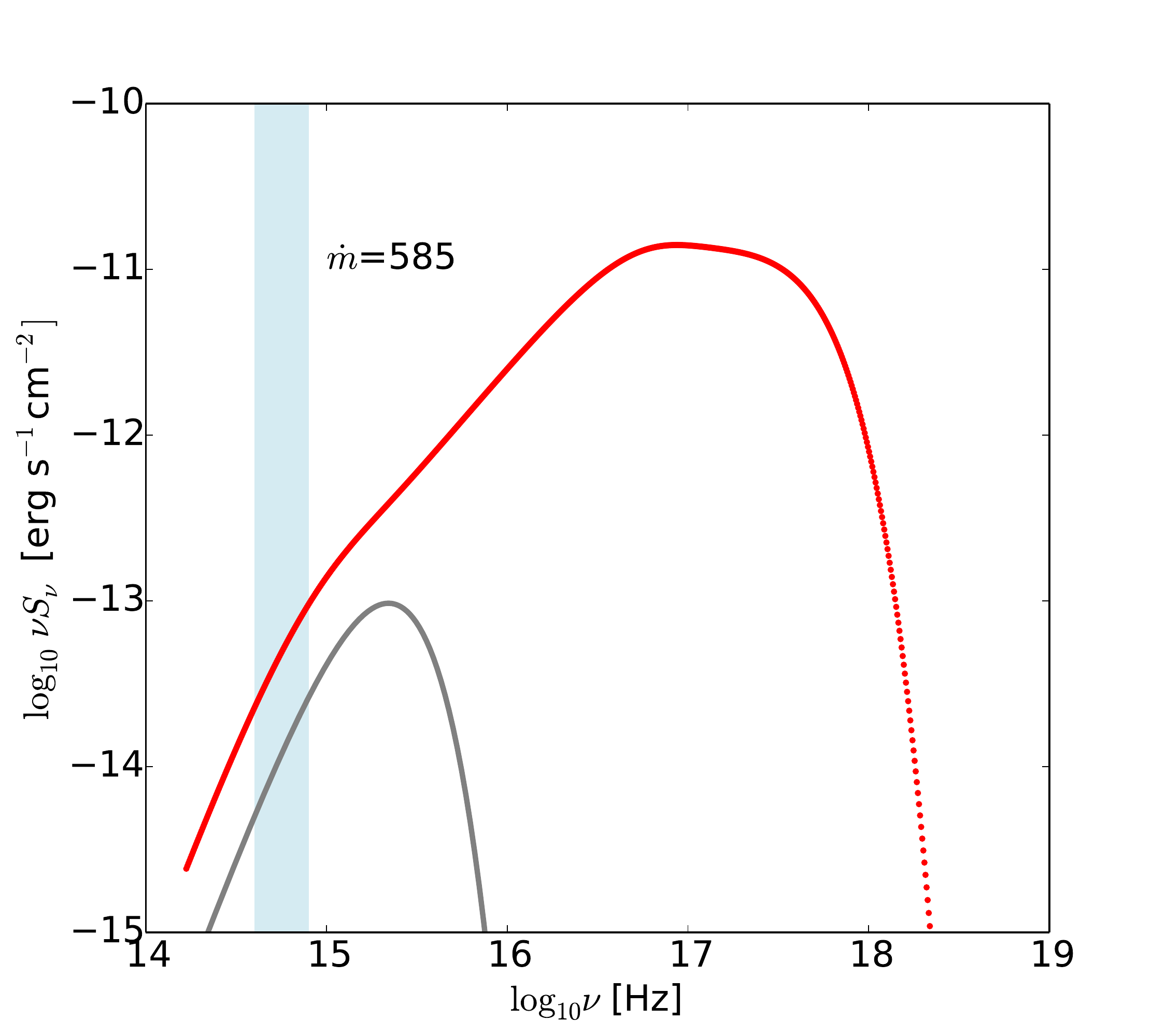}
\task
\includegraphics[width = 0.3\textwidth]{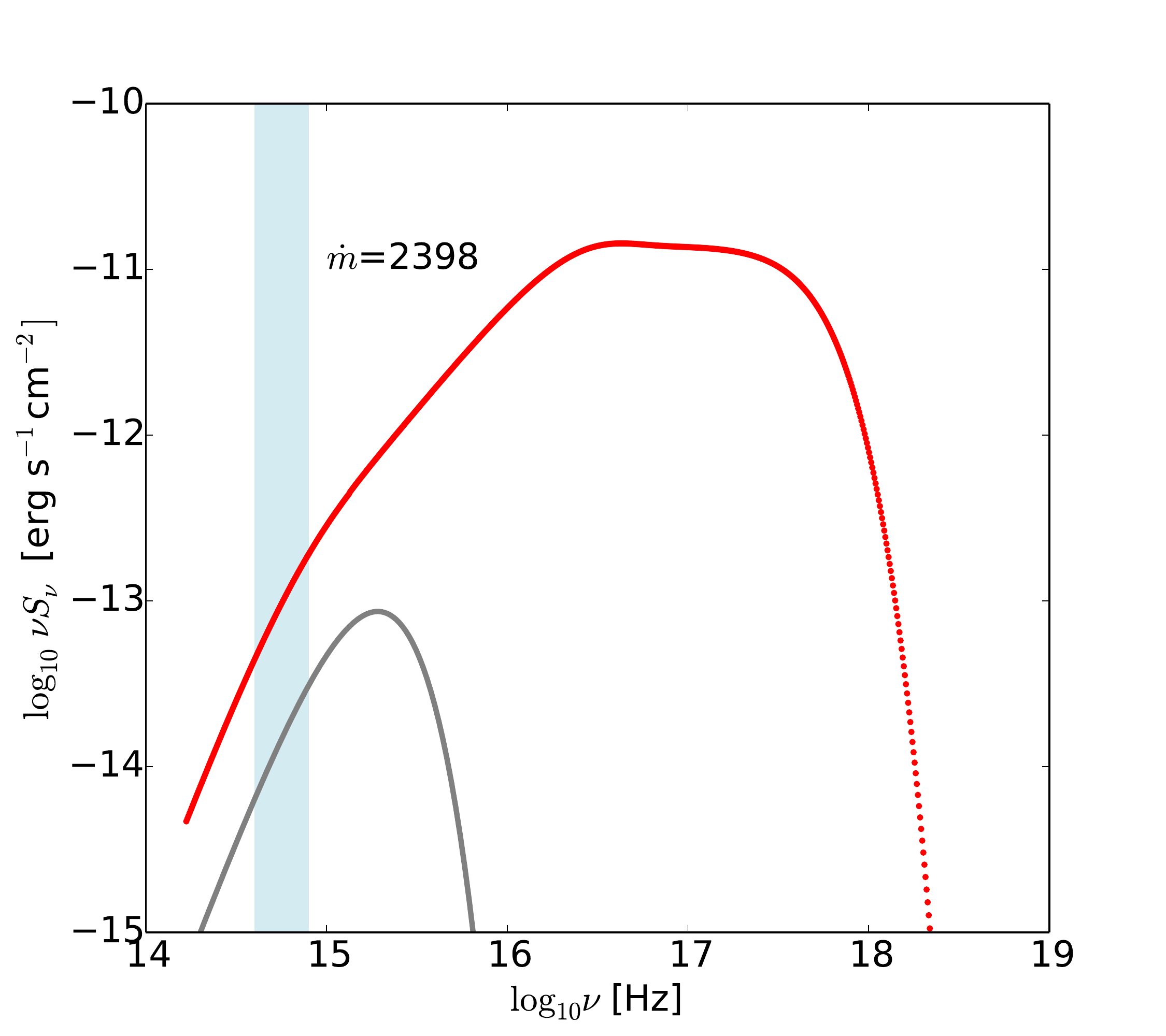}
\task
\includegraphics[width = 0.3\textwidth]{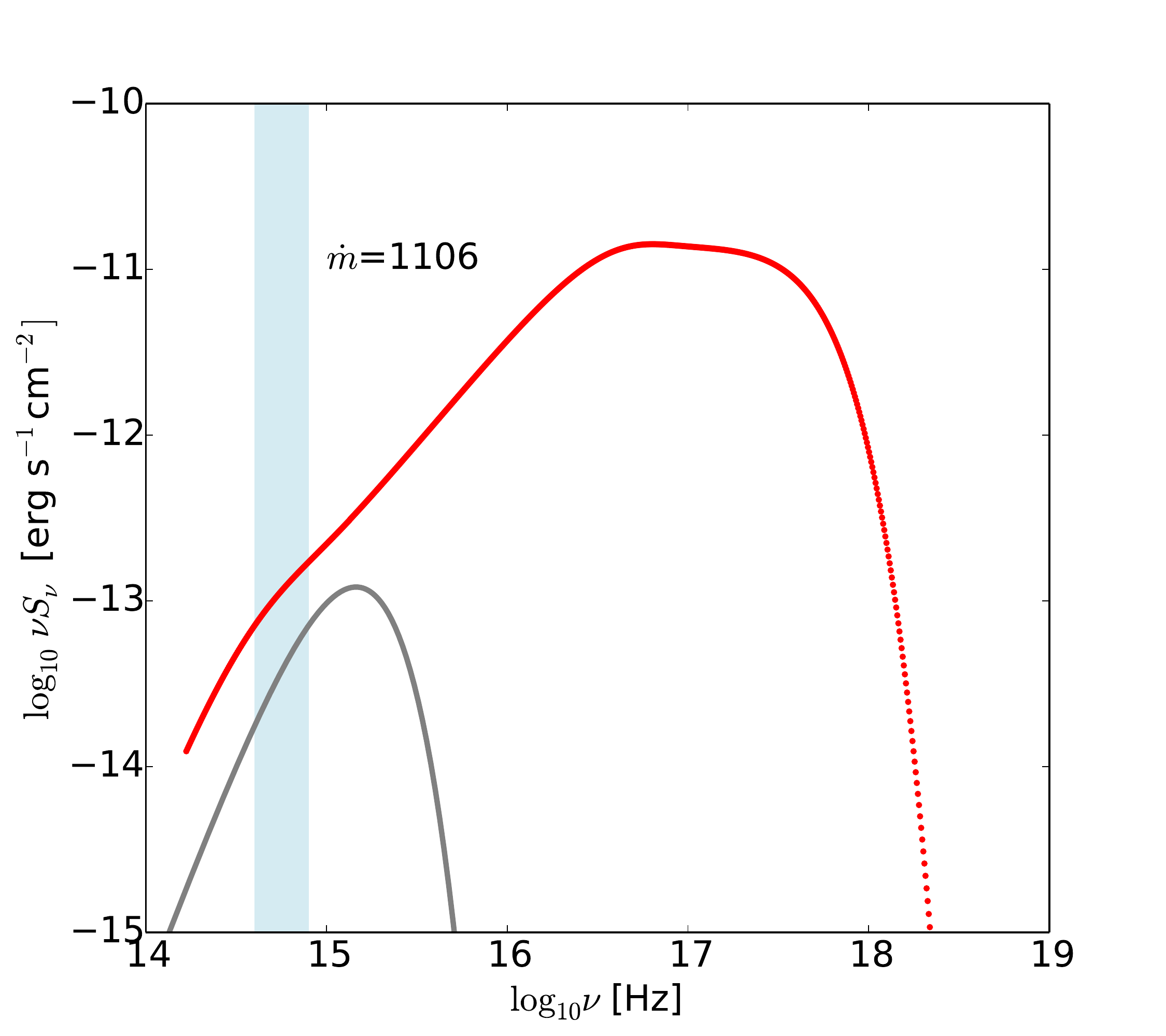}
\end{tasks}
\caption{Same as Figure~\ref{fig:spettro1020_42*} for a $25 M_{\odot}$ donor and a $100 M_{\odot}$ BH. The accretion rate is so high that the optical flux emitted from the disc dominates over that from irradiation. 
}
\label{fig:spettro25100_42*}
\end{figure*}
The synthethic spectra are computed for the same evolutionary phases considered before for the case with outflow ($a$, $b$ and $c$) and are shown in Figs.~\ref{fig:spettro1020_42*}-~\ref{fig:spettro25100_42*}.
When $\dot{m}$ is highly super-Eddington, the systems are more luminous than those of PZ. At the beginning of the Giant phase they are bluer while, at the final stages of their evolution (when the accretion disc is very extended), they become redder.


\bsp	
\label{lastpage}
\end{document}